\documentclass[10pt,titlepage]{article}
\usepackage{longtable}
\usepackage{array}
\usepackage{latexsym}
\usepackage{amssymb}
\usepackage{hyperref}
\usepackage{graphicx}
\usepackage{lscape}
\usepackage{color}

\newcommand{\be}{\begin{equation}}
\newcommand{\ee}{\end{equation}}
\newcommand{\bea}{\begin{eqnarray}}
\newcommand{\eea}{\end{eqnarray}}

\begin{document}

\pagestyle{empty}
\renewcommand{\thefootnote}{\alph{footnote}}
\begin{center}

{\Large \bf
Discovering the New Standard Model:\\ 
Fundamental Symmetries and Neutrinos
}
\vspace*{0.2in}

{\small 
Report of the Fundamental Symmetries and Neutrinos Workshop \\  
August 10-11, 2012  --  Chicago, IL\\
}
\vspace*{0.5in}

V.~Cianciolo,$^1$           
A.\,B.~Balantekin,$^2$    
A.~Bernstein,$^3$          
V.~Cirigliano,$^4$            
M.\,D.~Cooper,$^4$   
D.\,J.~Dean,$^1$        
S.\,R.~Elliott,$^4$        
B.\,W.~Filippone,$^5$     
S.\,J.~Freedman,$^{6,7,}$\footnotemark[1] 
G.\,L.~Greene,$^{8,1}$   
K.\,M.~Heeger,$^2$    
D.\,W.~Hertzog,$^{9}$   
B.\,R.~Holstein,$^{10}$    
P.~Huffman,$^{11,1}$    
T.~Ito,$^4$                     
K.~Kumar,$^{10}$     
Z.-T.~Lu,$^{12}$     
J.\,S.~Nico,$^{13}$     
G.\,D.~Orebi~Gann,$^{6,7}$     
K.~Paschke,$^{14}$                  
A.~Piepke,$^{15}$     
B.~Plaster,$^{16}$      
D.~Pocanic,$^{14}$     
A.\,W.\,P.~Poon,$^{7}$     
D.\,C.~Radford,$^{1}$     
M.\,J.~Ramsey-Musolf,$^{2}$     
R.\,G.\,H.~Robertson,$^{9}$     
G.~Savard,$^{12}$     
K.~Scholberg,$^{17}$     
Y.~Semertzidis,$^{18}$     
J.\,F.~Wilkerson$^{19,20,1}$   

\vspace*{0.2 in}

\textit{\scriptsize
$^1$\,Physics Division, Oak Ridge National Laboratory, Oak Ridge, TN\\
$^2$\,Department of Physics, University of Wisconsin, Madison, WI\\
$^3$\,Physics Division, Lawrence Livermore National Laboratory, Livermore, CA\\
$^4$\,Physics Division, Los Alamos National Laboratory, Los Alamos, NM\\
$^5$\,Department of Physics, California Institute of Technology, Pasadena, CA\\
$^6$\,Physics Department, University of California at Berkeley, Berkeley, CA\\
$^7$\,Nuclear Science Division, Lawrence Berkeley National Laboratory, Berkeley, CA\\
$^8$\,Department of Physics and Astronomy, University of Tennessee, Knoxville, TN\\
\mbox{$^{9}$\,Center for Experimental Nuclear Physics and Astrophysics, and Department of Physics, University of Washington, Seattle, WA}\\
$^{10}$\,Department of Physics, University of Massachusetts, Amherst, MA\\
$^{11}$\,Department of Physics, North Carolina State University, Raleigh NC\\
$^{12}$\,Physics Division, Argonne National Laboratory, Argonne, IL\\
$^{13}$\,Quantum Metrology Division, National Institutes of Standards and Technology, Gaithersburg, MD\\
$^{14}$\,Department of Physics, University of Virginia, Charlottesville, VA\\
$^{15}$\,Department of Physics and Astronomy, University of Alabama, Huntsville, AL\\
$^{16}$\,Department of Physics, University of Kentucky, Lexington, KY\\
$^{17}$\,Department of Physics, Duke University, Durham, NC\\
$^{18}$\,Physics Department, Brookhaven National Laboratory,  Upton, NY\\
$^{19}$\,Department of Physics and Astronomy, University of North Carolina, Chapel Hill, NC\\
$^{20}$\,Triangle Universities Nuclear Laboratory, Durham, NC\\
}

\vspace*{0.25in}

(\today)

\end{center}	

\footnotetext[1]{Deceased}
\renewcommand{\thefootnote}{\arabic{footnote}}

\newpage


\pagestyle{plain}
\renewcommand{\thepage}{\roman{page}}


\tableofcontents
\newpage

\listoftables
\addcontentsline{toc}{section}{List of Tables}
\newpage

\setcounter{page}{1}
\renewcommand{\thepage}{\arabic{page}}


\section{Executive Summary and Recommendations}
While the Standard Model (SM)  successfully describes a wealth of phenomena  over a wide range of energies, from atomic scales to hundreds of GeV, it has no answer to a number of questions about our universe: What is the  origin of  the  observed matter-antimatter  asymmetry?  How do neutrinos acquire mass?   What makes up dark matter? In addition to these  ``empirical" questions,  a number of theoretical puzzles  (such as the hierarchy of  weak and Planck scales, fermion family structure, unification)  indicate that  the SM is  incomplete, emerging most likely  as  the low-energy limit of a more fundamental theory  involving new degrees of freedom and interactions  --  the ``New Standard Model".  

The quest for the New Standard Model (NSM) is an interdisciplinary field of research where atomic, high-energy, astro- and nuclear physics communities meet, with nuclear physics playing a central role in many experimental and theoretical developments. High-sensitivity searches for rare/forbidden processes and exquisitely precise measurements of electro-weak nuclear processes are uniquely suited to
elucidate the NSM; allowing {\it discovery} of new sources of symmetry violation (B, L, CP,...) 
that are essentially inaccessible at the LHC, and helping to {\it discriminate} among various NSM models by
exploiting the fact that each model generates a unique pattern of  low-energy  signatures. 


As a result of current investments significant progress has been made in the last five years, leaving us poised for significant discoveries in the next decade. Examples include:  the only measurement of the correlation between neutron spin and its decay electron's momentum using ultracold neutrons (UCNA) -  together with new measurements of the neutron lifetime at NIST and abroad this gives a precise test of CKM unitarity; the most sensitive search for T-violation in $\beta$-decay (emiT); commissioning of NPDgamma; completion of  MuLan, TWIST, and MuCap with high precision measurements of the muon Fermi constant, Michel parameters, and weak-pseudoscalar couplings, respectively; measurement of  neutrino oscillation parameters $\theta_{12}$, $\Delta m^2_{12}$ (SNO and KamLAND) and $\theta_{13}$ (Daya Bay, Double CHOOZ and RENO); measurements of neutrino cross sections and searches for sterile neutrinos (MiniBooNE); new limits on neutrinoless double $\beta$-decay (EXO and CUORICINO); progress toward building the {\sc Majorana Demonstrator}, CUORE and KATRIN; and significant advances in parity-violation experiments at Jefferson Laboratory, with QWeak successfully completed, the PVDIS program well underway, and the MOLLER project, aiming for a new ultra-high precision measurement after the 12 GeV upgrade,  in the design phase.

In preparation for the upcoming NSAC subcommittee deliberations a workshop
\footnote{``Fundamental Symmetries and Neutrinos'', August 10-11, 2012. Linked talks and submitted two-page write-ups covering aspects of the subcommittee questions to the field can be found at \href{http://www.phy.ornl.gov/funsym}{http://www.phy.ornl.gov/funsym}.}
was held to gather and coordinate input from the scientific community working in this subfield. The outcome was a set of four recommendations that are crucial to achieving the community's scientific goals.

\paragraph{\underline {Recommendation 1:}} \textbf {Fundamental Symmetries and Neutrinos figure prominently in the 2007 NSAC Long Range Plan and subsequent assessments of the field, such as the 2012 NAS Nuclear Physics Report, the 2011 NAS Assessment of Science Proposed for DUSEL, and the 2011 NSAC Neutron subcommittee report. The community strongly endorses the recommendation of the 2007 NSAC Long Range Plan:  \textit {``We recommend a targeted program of experiments to investigate neutrino properties and fundamental symmetries. These experiments aim to discover the nature of the neutrino, yet-unseen violations of time-reversal symmetry, and other key ingredients of the New Standard Model of fundamental interactions. Construction of a Deep Underground Science and Engineering Laboratory is vital to U.S. leadership in core aspects of this initiative.''}}

The workshop made it clear that this subfield is vibrant and continues to address compelling scientific issues with considerable discovery potential. In particular, for the short to medium term timescales, the community strongly endorses neutrinoless double $\beta$-decay, neutron EDM, parity violating electron scattering, and the g-2 experiment. In addition, we note that progress has been made toward establishing an underground laboratory for science within the U.S.; at the Sanford Underground Research Facility, the Davis Campus at 4,850 feet is operating and being used as a home for the {\sc Majorana Demonstrator} as well as the LUX dark matter experiment. 

The targeted program of experiments to investigate neutrino properties and fundamental symmetries is envisioned to include next-generation experiments to measure neutrinoless double $\beta$-decay and the neutron electric dipole moment. Ton-scale double $\beta$-decay experiments will have sensitivity to the inverted hierarchy mass region and will improve our sensitivity to lepton-number violation by an order of magnitude. The U.S. nuclear physics community anticipates playing a leading role in selecting the most promising technologies and isotopes for these next generation double $\beta$-decay experiments, with support to start construction of at least one of these efforts. The proposed experiment to measure the neutron electric dipole moment will improve the current experimental limit by two orders of magnitude, offering unprecedented sensitivity to CP-violation in the quark sector.

It is important to recognize that this subfield is highly leveraged with facilities operations funded through offices other than Nuclear Physics. For example, the LANL LANSCE facility, operated by the NNSA, provides neutrons to the UCN source; the ORNL SNS, operated by Basic Energy Sciences, provides neutrons to the Fundamental Neutron Physics Beamline and a well-defined isotropic source of neutrinos; neutron beamlines at the NIST Center for Neutron Research, operated by the Department of Commerce, provide neutrons for a suite of fundamental neutron experiments; Sanford Underground Laboratory, with future funding from the state of South Dakota and DOE High-Energy Physics, provides space and infrastructure for the {\sc Majorana Demonstrator}; DOE's Waste Isolation Pilot Plant (WIPP) provides space for EXO. Furthermore, this community participates in several ongoing and relevant efforts such as the muon g-2 and Mu2e experiments that will be performed at Fermilab with a significant contribution expected from NSF and HEP. Finally, it is worth mentioning that the Continuous Electron Beam Accelerator Facility (CEBAF) at Jefferson Laboratory (JLab) provides a beam with characteristics that are uniquely suited for parity-violation experiments.

\paragraph{\underline{Recommendation 2:}} {\bf The federal research investment in Fundamental Symmetries and Neutrinos should be commensurate with its tremendous scientific opportunities and discovery potential.}  

The second recommendation follows the 2012 NAS Nuclear Physics Report in recognizing the need to balance research funding and construction of new facilities. Fundamental Symmetries and Neutrinos has traditionally not required a large facility operations budget, and therefore represents a cost effective way to obtain tremendously valuable science. However, substantial investments are critical for some of the next-generation experiments needed to maintain scientific momentum and world competitiveness. 

\paragraph{\underline{Recommendation 3:}} {\bf In order to ensure the long-term health of Fundamental Symmetries and Neutrinos research, it is necessary to establish and maintain a balance between funding construction of new experiments and facilities with the needs of university and laboratory-based research programs performing existing experiments and developing new ideas and measurements that may have high impact.}  

It is important to fund small R\&D efforts to establish the feasibility of new ideas. These ideas range from measuring sterile neutrino signals at the Spallation Neutron Source, to small-scale experiments at reactors to understand reactor neutrino flux predictions and spectra, and to measure salient features of neutrino oscillations and investigate coherent neutrino scattering, to initial R\&D towards a proton ring for measuring the proton electric dipole moment and using novel polarized photon sources to investigate parity violation in both light and heavy nuclei. 

\paragraph{\underline{Recommendation 4:}} {\bf The community urges strengthened support for nuclear theory in Fundamental Symmetries and Neutrinos in order to fully exploit, guide, and complement experimental efforts. } 


Although relatively few in number, nuclear theorists from both universities and national laboratories have historically played a crucial role in developing  Fundamental Symmetries and Neutrinos. Their efforts have been especially important in connecting the results of different experiments and providing an intellectual bridge to astrophysics and particle physics. It is crucial to strengthen these efforts to capture the opportunities this field represents. In addition, training the next generation of nuclear theorists starts at universities and we strongly encourage agency incentives to universities (for example, bridge appointments) in this subfield of nuclear theory.


\clearpage
\section{Scientific Landscape}

The quest to understand nature's fundamental interactions, and how they have shaped the evolution of the cosmos, is among the most compelling in modern science. The Standard Model - a comprehensive and detailed picture of the electroweak and strong interactions - is a triumph for that quest. And yet, at best it is incomplete, emerging most likely  as  the low-energy limit of a more fundamental theory involving new degrees of freedom and interactions  --  dubbed the ``New Standard Model" (NSM) in the 2007 NSAC Long Range Plan (LRP)~\cite{LRP2007}.  

The quest for the NSM is an interdisciplinary field of research where atomic, high-energy, astro- and nuclear physics communities meet, with nuclear physics playing a central role in many experimental and theoretical developments. At the high-energy frontier new particles in the few-TeV mass range can be directly excited, and the LHC discovery of a SM-like Higgs particle~\cite{HiggsCMS, HiggsATLAS} has opened a new era in electroweak symmetry breaking studies. This White Paper focuses on the recent progress and future outlook  at the low-energy precision frontier where the U.S. Nuclear Physics community plays a key leadership role.

The 2007 NSAC LRP laid out a targeted series of experiments, a ``New Standard Model Initiative",  with the goal of answering three overarching questions:
\begin{itemize}
\item {\it ``What is the nature of the neutrinos, what are their masses, and how have they shaped the evolution of the cosmos?"}
\item {\it ``Why is there now more visible matter than antimatter in the universe?"}
\item {\it ``What are the unseen forces that were present at the dawn of the universe but disappeared from view as it evolved?"}
\end{itemize}
These experiments can be divided into two broad classes, ({\it i}) high-sensitivity searches for rare/forbidden processes, and ({\it ii}) exquisitely precise measurements of electro-weak nuclear processes that serve to establish precise values for SM parameters and search for subtle violations of SM predictions. These experiments are uniquely suited to elucidate the NSM, allowing {\it discovery} of new sources of symmetry violation (B, L, CP,...) that are essentially inaccessible at the LHC, and helping to {\it discriminate} among various NSM models by exploiting the fact that each model generates a unique pattern of low-energy signatures. 

Work at both the high-energy and low-energy precision frontiers is needed to reconstruct the structure, symmetries, and parameters of the underlying New Standard Model. With new physics signals possibly arising from the LHC,  low-energy probes at the current/planned level of precision will be essential to understand the NSM symmetries and diagnose the underlying dynamics. One could argue that the motivation for low-energy precision frontier experiments will be even stronger if there are no clear NSM signals at the LHC: in that case a broad set of searches with mass reach above 10 TeV will be needed to go beyond the Standard Model since it may not be clear what to look for or where.  Recent investments have led to significant progress on a number of low-energy precision experiments, leaving us poised for significant discoveries in the next decade.


\clearpage
\section{Physics Beyond the Standard Model: New Phenomena}
Highly suppressed or strictly forbidden in the  SM  due to approximate or exact symmetries,  observables in this category include neutrinoless double $\beta$-decay of nuclei ($0\nu\beta\beta$), permanent electric dipole moments (EDMs) of leptons, nucleons, and atoms, and charged lepton flavor violating  processes  such as  $\mu \to e$ conversion in nuclei. These processes typically probe very high-energy scales, above the reach of the LHC,  and   a positive signal in such  observables would provide ``smoking gun" evidence for  physics beyond the Standard Model. Among these processes, a special role is played by $0\nu\beta\beta$ and EDM searches that shed light on various mechanisms for  the origin of the baryon asymmetry of the universe (BAU). In fact, $0\nu\beta\beta$ searches  probe total lepton number violation, which is a key ingredient in leptogenesis scenarios for the  BAU generation. On the other hand,  EDM searches  probe new sources of CP violation (which is also a key ingredient to understand  the BAU),  and can strongly  constrain weak-scale  baryogenesis scenarios. 

\subsection{Matter Asymmetry of the Universe}
The matter asymmetry of the universe remains one of the deepest mysteries in physics.  The 
absence of significant amounts of antimatter requires, as Sakharov~\cite{Sakharov} explained, a time when 
the universe was not in equilibrium, the non-conservation of baryon and lepton number, and 
violation of CP invariance.  Non-conservation of baryon and lepton number has not been 
experimentally discovered, despite heroic efforts to observe proton decay.  Violation of 
CP in the quark sector is well established but  insufficient to explain the asymmetry.  
As a consequence, it is essential to explore new sources of CP violation, and searches for 
static electric dipole moments and  neutrinoless double $\beta$-decay are high priorities in 
nuclear physics.  The observation of neutrinoless double $\beta$-decay would  further demonstrate the 
non-conservation of lepton number, and by inference the non-conservation of baryon number.

\subsubsection* {Neutrinoless Double $\beta$-Decay}

The search for $0\nu\beta\beta$ decay is a primary goal of the NP neutrino program, with the 
potential to determine the fundamental nature of the neutrino itself, and shed light on 
questions such as the source of the matter-antimatter asymmetry in our universe.
In contrast to all other fundamental building blocks of matter neutrinos carry no 
electrical charge, opening the possibility that they are their own antiparticles if no 
conserved quantum number forbids it.  The experimental demonstration of neutrino oscillations 
shows that neutrinos have mass, and therefore that neutrinoless double $\beta$-decay can occur if neutrinos are their own antiparticles, i.e. Majorana neutrinos.  

The rate of neutrinoless double $\beta$-decay ($0\nu\beta\beta$) can be written as:   
\begin{eqnarray}
\left[T_{1/2}^{0\nu}\right]^{-1} 
&=&G_{0\nu}\cdot g_A^4\cdot \left|\frac{\langle m_{ee} \rangle}{m_e}\right|^2\cdot \left|M_{0\nu}\right|^2
\label{eq:half_life_me}
\end{eqnarray}
where $G_{0\nu}$ is the phase space factor, $M_{0\nu}$ is the nuclear matrix element, and 
\begin{eqnarray}
\langle m_{ee}\rangle &=& \left|U_{e1}^2\cdot m_1+U_{e2}^2\cdot m_2\cdot e^{i\alpha}+U_{e3}^2\cdot m_3\cdot 
e^{i\beta}\right|.
\label{eq:m_ee}
\end{eqnarray}
is the effective Majorana mass, a coherent sum over mass eigenstates with (potentially) CP-violating phases
in which cancellations can occur.  $\langle m_{ee} \rangle$ can also be modified by interference with other 
hypothesized non-Standard Model processes. 

Recent publications~\cite{Rod03, Rod06, Suh05, Kor07, Fae08, Sim09a, Sim09} have eliminated the historical disparity of results from quasiparticle random phase approximate (QRPA) calculations of $M_{0\nu}$~\cite{Sim09}. The Nuclear Shell Model (NSM) has also seen a resurgence of activity in recent years with studies of input micro-physics and its influence on $M_{0\nu}$ ~\cite{Cau08, Cau08a, Men09}. Three additional techniques have recently been used to look at $M_{0\nu}$; the Interacting Boson Model (IBM-2)~\cite{Bare09}, the Projected Hartree-Fock-Bogoliubov (PHFB)~\cite{Chan09}, and the Energy Density Functional (EDF)~\cite{Rod10} techniques. The two QRPA groups -- Tuebingen-Caltech-Bratislava\cite{Sim09a} and Jyv\"{a}skyl\"{a}-La Plata~\cite{Suh08} -- and the NSM~\cite{Men09a} group each recalculated their $^{76}$Ge matrix elements requiring that they reproduce new measurements of $^{76}$Ge and $^{76}$Se occupation numbers~\cite{Schif08,Kay09}, resulting in significantly improved agreement between the different techniques.

Experimental progress in double $\beta$-decay has been rapid since 2007.
As summarized in Table~\ref{tab:0nubb_experiments}, a number of experiments - using a variety of experimental techniques and isotopes - are presently beginning operations or under construction. This current generation of experiments aims to reach $\langle m_{ee}\rangle <100$\,meV. This is an order of magnitude below the current laboratory limits for the kinematic mass (2.2\,eV). Specifically, a kinematic mass of 2.2\,eV corresponds to an effective Majorana mass between 2.2 and 0.7 eV, depending on the phases. These measurements will also definitively address the existing $0\nu\beta\beta$ decay claim~\cite{kla06}.

The combination of neutrino oscillation data and the light neutrino exchange dominance hypothesis allows the computation of allowed $\langle m_{ee}\rangle$ ranges for the degenerate, inverse-hierarchy, and normal-hierarchy neutrino mass scenarios. Oscillation data indicate that double $\beta$-decay searches with mass sensitivities of about 50\,meV and 15\,meV are needed to cover the degenerate or inverse-hierarchy scenarios, respectively~\cite{rodejohann_2012}. Because of the possibility of destructive interference there is no lower limit in the normal hierarchy, and therefore limit results do not by themselves yield definitive information on the mass ordering and Majorana or Dirac character of neutrinos (although they could do so if independent information about the kinematic mass becomes available from experiments such as KATRIN, or from cosmology). If no signal is seen with the current generation of experiments their results will be used to optimize the strategy for next-generation experiments which would need significantly larger (ton-scale) quantities of enriched isotope to have some chance for a sufficient signal to be detected.

\begin{landscape}

\hfill

\begin{table}
\caption{Summary of current generation $0\nu\beta\beta$ decay experiments.}
\label{tab:0nubb_experiments}
\begin{minipage}{9.0in}
\begin{center}

\begin{tabular}{|c|l|c|c|c|l|}\hline
{\bf Experiment} & {\bf Technology} & {\bf Location} & {\bf Active}   & {\bf Active Isotope} & {\bf Notes} \\
                 &                  &                & {\bf Isotope}  & {\bf Fiducial Mass}  &             \\ \hline\hline
                         & Tracking calorimeter w/      &&&                             & Energy resolution: $\sigma/Q$ = 1.67\% \\
EXO-200~\cite{Auger2012} & simultaneous readout of      & WIPP & $^{136}$Xe & 585  mole & Current result: $\langle m_{ee}\rangle  <140-380$~meV  \\
                         & ionization and scintillation &&&                             & \\ \hline
{\sc Majorana }          & High-resolution Ge           &&&                             & Under construction, first data in 2013\\
{\sc  Demonstrator}      & semiconductor detectors      & SURF & $^{76}$Ge  & 340  mole & Projected energy resolution: $\sigma/Q$ = 0.1\% \\
(MJD)~\cite{Aguayo2011,Phillips2011,Schubert2011} &     &&&                             & Projected sensitivity: 
                                                                                          $\langle m_{ee}\rangle < 100-210$~meV\\ \hline
&&&&                                                                                     & Under construction, first data in 2014\\
CUORE~\cite{Aless11,And11} & Cryogenic bolometers & Gran Sasso & $^{130}$Te & 1600  mole & Prototype result: $\langle m_{ee}\rangle  <300-700$~meV \\
&&&&                                                                                     & Projected sensitivity: 
                                                                                           $\langle m_{ee}\rangle < 40-90$~meV\\ \hline
&&&&                                                                                     & Current result: $\langle m_{ee}\rangle <128-348$~meV \footnote{Derived from quoted half-life, $\tau_{1/2} > 1.9 \times 10^{25}$\,y (90\% C.L.U.L), using the same nuclear matrix elements as the EXO-200 result.}\\
KamLAND-Zen                & Liquid scintillator  & Kamioka    & $^{136}$Xe & 1316  mole & 500 kg enriched $^{136}$Xe available for \\
\cite{Gando2012, KLZ_new}  &                      &&&                                    & \hspace*{0.15in} future upgrades \\ 
&&&&                                                                                     & Energy resolution: $\sigma/Q$ = 4.2\% \\ \hline
&&&&                                                                                     & Under construction, first data in 2014\\
SNO+~\cite{che05,Chen2008} & Liquid scintillator & SNOLab      & $^{150}$Nd & 440  mole  & Projected sensitivity: 
                                                                                           $\langle m_{ee}\rangle < 90-175$~meV\\
&&&&                                                                                     &  \\ \hline
&&&&                                                                                              & Quoted fiducial mass (moles) for 10\,atm scenario \\
NEXT~\cite{Gom11,Yah10}    & High-pressure TPC   & LSC         & $^{136}$Xe & $\approx 400$ mole  & Assembly and commissioning in 2014 \\
&&&&                                                                                              & First data in 2015 \\ 
&&&&                                                                                              & Projected sensitivity:
                                                                                                    $\langle m_{ee}\rangle < 80-130$~meV
\footnote{Assumes three years of running, 10\,atm scenario, and energy resolution scaling as $1/\sqrt{E}$ from calibrations at 662\,keV.} \\ \hline
&&&&                                                                                              & \\
Super-NEMO                 & Tracking calorimeter & Fr\'{e}jus & $^{82}$Se  & $\approx 85$ mole   & Commissioning in 2013 -- 2014\\
\cite{Arn10, SUPERNEMO_new}&&&&                                                                   & Projected sensitivity:
                                                                                                    $\langle m_{ee}\rangle < 200-400$~meV\\ \hline
\end{tabular}
\end{center}
\end{minipage}
\end{table}

\end{landscape}

\subsubsection* {Electric Dipole Moments}

A permanent EDM is a sensitive probe for new CP-violating mechanisms, and is generally considered to be one of the most promising paths towards new physics beyond the Standard Model. The CKM mechanism can only generate EDMs at the three- and four- loop level, leading to values many orders of magnitude below current experimental limits. Therefore, any observed non-zero EDM would require either CP-violation in the strong interaction or physics beyond the Standard Model. The 2011 Fundamental Neutron Physics NSAC (Kumar) report~\cite{Kumar2011} reiterated the scientific motivation for EDM searches, saying they remain as compelling as ever. This search will challenge theories for physics beyond the Standard Model and the weak baryogenesis hypothesis regarding the baryon asymmetry of the universe.

Since the last long range plan, the most important EDM results have been the new limit for $^{199}$Hg  ($0.3 \times 10^{-28}$~e$\cdot$cm~\cite{Griffith09}), for the neutron ($300 \times 10^{-28}$~e$\cdot$cm~\cite{Baker06}), and for the electron ($10 \times 10^{-28}$~e$\cdot$cm~\cite{Hudson11}).  These measurements are complementary in that they are each sensitive to different possible sources of CP-violation.

\paragraph{Atomic EDMs} The $^{199}$Hg measurement will be pursued with a goal of another factor of ten improvement in precision in the next five years, and an additional factor of five after 2020.  In order to take advantage of significant enhancements of the Schiff moment in nuclei with permanent octupole deformation, promising experiments on $^{225}$Ra~\cite{De09, Guest07} and $^{221,223}$Rn are underway that should produce exciting results after 2020. Due to the short half-lives of these nuclear species, both experiments would benefit from next generation isotope production facilities.

\paragraph{Proton EDM} The storage ring EDM collaboration has submitted  a proposal to DOE NP  for a proton EDM experiment sensitive to $0.1 \times 10^{-28}$ e$\cdot$cm~\cite{edmweb}.  The method utilizes polarized protons in an all-electric storage ring at the so-called ``magic'' momentum of 0.7 GeV/{\it c} where the proton spin and momentum vectors precess at the same rate in any transverse electric field, allowing the radial E-field act on the proton EDM and precess its spin out of the horizontal plane for the duration of the storage time (1,000\,s). Systematic errors due to efficiency and analyzing power of the polarimeter and the geometric phase have been shown to be lower than the statistical sensitivity.

\paragraph{Neutron EDM} Eight neutron EDM experiments, summarized in Table~\ref{tab:nedm1}, have begun worldwide. At least one of these should produce a factor of five sensitivity improvement by 2020. The Kumar report~\cite{Kumar2011} stated that a measurement of the neutron EDM with sensitivity at the ultimate reach of the nEDM@SNS experiment (the primary U.S. effort) would have a profound impact on nuclear physics, particle physics and cosmology, even in the event of a negative result, and deemed this to be the initiative with the highest scientific priority in U.S. neutron science.    

\begin{table}[h]  
    \begin{center}
    \caption{Summary of worldwide nEDM searches.}
    \label{tab:nedm1}
    \begin{tabular}{|l|l|l|l|l|}
        \hline
        \multicolumn{1}{|c|}{\textbf{Experiment}} & \multicolumn{1}{c|}{\textbf{UCN Source}} & 
	\multicolumn{1}{c|}{\textbf{Cell}} & \multicolumn{1}{c|}{\textbf{Measurement}} & 
	\multicolumn{1}{c|}{$\mathbf{\sigma_{d}}$} \\
         &  &  & \multicolumn{1}{c|}{\textbf{Technique}} & \multicolumn{1}{c|}{\textbf{($\mathbf{10^{-28}}$ e$\cdot$cm)}} \\
        \hline
        CryoEDM (ILL)& Superfluid $^{4}$He & $^{4}$He & Ramsey technique for $\omega$     & Phase 1 $\approx 50$   \\
           &                     &          & External SQUID magnetometers      & Phase 2 $< 5$          \\
        \hline
        PNPI (ILL)    & ILL turbine         & Vacuum   & Ramsey technique for $\omega$     & Phase 1 $< 100$        \\
          & PNPI/Solid D$_{2}$  &          & $\vec{E}=0$ cell for magnetometer & Phase 2 $< 10$         \\
        \hline
        Crystal (ILL) & Cold neutrons       & Solid    & Crystal Diffraction               & $< 100$  \\
        \hline
        PSI EDM & Solid D$_{2}$       & Vacuum   & Ramsey technique for $\omega$     & Phase 1 $\approx 50$   \\
                &                     &          & External Cs and $^{3}\vec{\mbox{He}}$ magnetometers      & Phase 2 $< 5$          \\
         &  &  & Possible Hg or Xe comagnetometer &   \\
        \hline
        Munich FRMII & Solid D$_{2}$  & Vacuum   & Under construction                & $< 5$   \\
                &                     &          & Similar to PSI EDM                &           \\
        \hline
        nEDM  (SNS)   & Superfluid $^{4}$He & $^{4}$He & $^{3}$He capture for $\omega$     & $< 5$   \\
          &                     &          & $^{3}$He comagnetometer           &           \\
                &                     &          & \textsc{squids} \& Dressed spins  &   \\
        \hline
        TRIUMF  & Superfluid $^{4}$He & Vacuum   & Phase I \@ RCNP                   & $< 10$   \\
        \hline
        JPARC   & Solid D$_{2}$       & Vacuum   & Under development                 & $< 5$   \\
        \hline
     \end{tabular}
    \end{center}
\end{table}

The nEDM@SNS experiment, based on Ref.~\cite{GolubLamoreaux}, uses a novel polarized $^3$He co-magnetometer and will detect the neutron precession via the spin-dependent neutron capture on $^3$He. High densities of trapped ultra-cold neutrons are produced via phonon production in superfluid $^4$He which can also support large electric fields. The capture signal is observed via scintillation light produced from the ionization in liquid helium of the energetic proton and triton produced in the reaction.  A value (or limit) for the neutron EDM will be extracted from the difference between neutron spin precession frequencies for parallel and anti-parallel magnetic ($\sim 30$ mG) and electric ($\sim 70$ kV/cm) fields. Control of systematic errors is essential for an experiment at the $10^{-28}$~e$\cdot$cm level.  Teams leading the different experiments have chosen different approaches, but the nEDM@SNS experiment has the most extensive program for estimating systematic errors, see Table~\ref{tab:nedm2}.

The nEDM@SNS collaboration, including researchers from twenty-one institutions with expertise in nuclear, atomic, and low-temperature physics, is continuing to address critical R\&D developments in preparation for construction of a full experiment, which will be carried out on the ORNL/SNS Fundamental Neutron Physics Beamline. Construction is likely to take at least five years, followed by hardware commissioning and data taking. Thus first results could be anticipated by the end of the decade.

\begin{table}[h]  
    \begin{center}
    \caption[Comparison of capabilities for different nEDM experiments.]{Comparison of capabilities for different nEDM experiments.  Items marked with a ``$^{*}$" indicate an advantage in addressing systematic errors if present.}
    \label{tab:nedm2}
\begin{tabular}{|l|c|c|c|c|c|}
    \hline
    \multicolumn{1}{|c|}{\textbf{Capability}} & \textbf{Cryo1} & \textbf{Cryo2} & \textbf{PSI2} & \textbf{PSI3} & \textbf{SNS}  \\
    \hline
    $\Delta\omega$ via accumulated phase in n polarization   & Y & Y & Y & Y & N  \\
    \hline
    $\Delta\omega$ via light oscillation in $^{3}$He capture & N & N & N & N & Y  \\
    \hline
    $^{*}$Comagnetometer                                     & N & N & Y & Y & Y  \\
    \hline
    $^{*}$Superconducting B-shield                           & Y & Y & N & N & Y  \\
    \hline
    $^{*}$Dressed Spin Technique                             & N & N & N & N & Y  \\
    \hline
    Horizontal B-field                                       & Y & Y & N & N & Y  \\
    \hline
    $^{*}$Multiple EDM cells                                 & N & Y & N & Y & Y  \\
    \hline
    $^{*}$Temperature Dependence of Geo-phase effect         & N & N & N & N & Y  \\
    \hline
\end{tabular}
    \end{center}
\end{table}

\subsection{Large-scale Structure, Origin of Neutrino Mass}
Neutrino flavor oscillations revealed that neutrinos have mass, a clear departure from the Standard Model. Since then the oscillation parameters (mixing angles and  mass differences) have been determined, but questions remain - What is the absolute neutrino mass scale? Are there sterile neutrinos?  

\subsubsection*{Neutrino Oscillations}

Neutrino oscillations have been measured using a variety of neutrino sources including atmospheric, solar, reactor, and accelerator based experiments. From the resolution of the solar neutrino problem at SNO to the discovery of reactor antineutrino disappearance with KamLAND, nuclear physics-funded experiments have played an important role in the history of neutrino physics and have made precision measurements of several neutrino oscillation parameters. Solar experiments are primarily sensitive to $\theta_{12}$ and SNO's measurement of $\theta_{12}$ provides the most precisely known neutrino mixing angle, which is pivotal both in testing various theories of neutrino symmetries and also for $0\nu\beta\beta$ experiments, since knowledge of this parameter defines the lower bound of the inverted hierarchy region.

KamLAND discovered reactor antineutrino disappearance and was the first experiment to observe the explicit $L/E$-dependence predicted by oscillations in the solar $\Delta m^2$ range~\cite{Abe08}, thus establishing oscillations as the solution to the flavor change observed by SNO. The final KamLAND results produced a precision ($< 3\%$) measurement of $\Delta m^2_{12}$.  In combination with solar experiments, this allowed determination of the mass ordering of the $\nu_1$ and $\nu_2$ states; a similar measurement in the 1 -- 3 sector would be a huge step in this field. KamLAND is currently collecting zero-reactor-power data, which will allow a direct measurement of detector background.  A new generation of reactor experiments including Double CHOOZ~\cite{Matsubara2012}, Daya Bay~\cite{An2012},  and RENO~\cite{Ahn2012} has recently measured $\theta_{13}$ and completed our understanding of the neutrino mixing matrix in the 3-neutrino framework. A new project called Daya Bay-II (but independent from the original Daya Bay experiment) is investigating the possibility of using a sub-\%-level precision reactor measurement, utilizing a 20~kton liquid scintillation detector to be located about 60 km from the Daya Bay reactors, to determine the mass hierarchy. SNO+ will also be sensitive to reactor neutrinos, with a lower flux than KamLAND but a cleaner spectrum, due to the very similar baseline of the small number of nearby reactors, thus allowing improved precision on the oscillation parameters in combination with existing data.

A recent re-analysis of short-baseline reactor neutrino experiments has revealed a discrepancy of 6\% between observations and the predicted antineutrino flux. This reactor anomaly
can be interpreted as a sign of neutrino oscillation with $\Delta m^2 \approx 1$\,eV$^2$ or could be due to an unknown issues with the reactor antineutrino flux predictions. Current km-scale reactor experiments, while highly precise, cannot probe such short oscillation lengths. The U.S. operates world-leading compact research reactors such as NIST~\cite{NISTNCR}, HFIR~\cite{HFIR}, and ATR~\cite{ATR} that provide access to very short baselines of $<10$\,m and scientific user support. The facilities and experience in the U.S. provide a unique opportunity for a world-leading, reactor experiment at the shortest baselines ever probed that can resolve the reactor anomaly, provide a precision measurement of the reactor anti-$\nu$ spectrum and test the hypothesis of sterile neutrinos. By deploying a high-strength radioactive source -- such as $^{51}$Cr or $^{144}$Ce -- Borexino, DayaBay, KamLAND and SNO+ could also test the recent short-baseline reactor anomaly.  

The MiniBooNE experiment has produced results in short-oscillation baseline accelerator neutrinos~\cite{Aguilar2009} and anti-neutrinos~\cite{Aguilar2010}. These data are consistent with the excess previously seen by the LSND measurement~\cite{agu01} and can be interpreted as an oscillation signal involving sterile neutrinos with  $\Delta m^2$ values near 1\,eV$^2$. These results have been considered along with other previous anomalous results to better understand constraints on sterile neutrinos~\cite{Abazajian2012}. It is clear that the rich phenomenology of neutrino mixing still permits such particles to exist. The MiniBooNE collaboration recently submitted a letter of intent to the FNAL PAC to add scintillator to their detector which would allow neutral-current gamma events to be distinguished from charged-current electron events. The collaboration also plans to request more running in neutrino mode with the goal of doubling the neutrino data sample. A joint analysis of MiniBOONE data with that from the approved new MicroBOONE experiment (a 170T liquid argon TPC along the booster beam line) should reduce the overall systematic uncertainties.

The ORNL Spallation Neutron Source (SNS) produces an intense flux of neutrinos (from $\pi^+$ and $\mu^+$ that decay at rest) with small spectral and flavor-content uncertainties and a 10$^{-3}$-10$^{-4}$ background rejection factor due to the short-pulse time structure~\cite{Avignone:2003ep}. This provides an outstanding opportunity to search for short baseline neutrino oscillations and sterile neutrinos~\cite{nusns, snsworkshop}. A kiloton size detector filled with dilute scintillator oil and covered by a few thousand 8-inch PMTs would have the capability of measuring neutrino oscillations in the detector with precision and proving the existence of sterile neutrinos if they exist at the eV mass scale~\cite{Efremenko:2008an}. 

\subsubsection*{Neutrino Mass}

Complementary to neutrinoless double $\beta$-decay, the shape of the electron spectrum in nuclear $\beta$-decay provides direct information about the neutrino mass, independent of the nature of the neutrino. Significant progress has been made on upcoming experiments to determine the neutrino mass. KATRIN is under construction in Germany, with significant U.S. nuclear physics participation.  Using magnetic collimation of electrons from a very strong tritium $\beta$-source, and a large electrostatic spectrometer, KATRIN will achieve a sensitivity of 0.2~eV, an order of magnitude better than current limits~\cite{Robertson2008}.  Data taking is planned to start in 2015. A new technology developed for the Project 8 experiment~\cite{Monreal2009} uses cyclotron radiation to measure $\beta$ energies and may extend past the limits of KATRIN. Other efforts using cryogenic bolometers, including ECHO~\cite{Gatti2008} and MARE~\cite{Andreotti2007}, are also progressing. In a complementary approach, the Planck satellite was launched in 2009 and upcoming observations of the cosmic microwave background and large-scale structure in the universe should lead to a sensitivity below 100 meV for the sum of the neutrino masses~\cite{Hannestad2006,Lesgourgues2006}. 

\subsection{Symmetries and Fields of the New Standard Model}
A complete understanding of the New Standard Model will require reconstructing its field content and symmetry structure. Low-energy experiments  can indirectly probe the existence of new heavy states, but can also directly excite new light and weakly coupled particles, such as a hypothetical ``dark photon", that can be searched for at JLab. Moreover, low-energy searches are uniquely sensitive to the flavor symmetries of the New Standard Model.  Particularly exciting are  prospects in the lepton flavor sector. Here,  while the observation of neutrino flavor oscillations has proved that individual lepton family number is not conserved, the search for  lepton flavor violation in charged lepton processes involving  electrons, muons and taus is central in uncovering the flavor structure  of the New Standard Model. In fact, given the extreme suppression of lepton flavor violating processes in the Standard Model (with BRs at the $10^{-54}$ level),  these searches have a great discovery/constraining  potential~\cite{Cirigliano:2009bz}: if the flavor symmetries are maximally broken, current and future experiments will probe very high mass-scales,  of the order of $10^4$~TeV. Should new physics originate at the TeV scale and have moderate flavor breaking, positive signals are expected in the next generation of $\mu \to e$ experiments.

\subsubsection*{Mu2e} 
Mu2e~\cite{mu2e} is a funded (HEP) experiment that will be sited at the Muon Campus under construction at Fermilab. This very ambitious experiment, has a single event sensitivity in the (essentially forbidden) charged lepton violation reaction $\mu \rightarrow$~e of $2 \times 10^{-17}$.  The experimental concept of the Mu2e design strategy has a long history, with a false start in the MECO effort at BNL. However, from that effort, a tremendous amount of design work has laid the foundation for the current experiment.  One of the many challenging features is a series of three connected superconducting solenoids designed for muon ``production'', ``transport'' and ``decay.''  Accelerator challenges include the need for a nearly $10^{10}$ suppression of protons between the $1.6~\mu$s bursts of protons that enter the production solenoid and create pions.  On the detector side, a very thin and high-rate straw tracker must identify the unique 105\,MeV decay signature against decay-in-orbit backgrounds and other backgrounds arising from ordinary muon capture.  The experiment involves a mix of particle-, nuclear- and accelerator-physicists with nuclear physicists providing expertise in ordinary muon capture, detectors and electronics.

\subsubsection*{Dark Photons}
A ``dark photon'' is the massive mediator of a new Abelian force.  It has a small coupling to electrically charged particles induced 
by ``kinetic mixing''~\cite{Holdom:1985ag,Galison:1983pa,Fayet:2007ua,ArkaniHamed:2008qp,Goodsell:2009xc,Cheung:2009qd,Morrissey:2009ur,Jaeckel:2010ni} with the hypercharge gauge boson.  The existence of dark photons is an attractive idea theoretically, and has recently received renewed attention due to the possibility that they could explain the discrepancy between the measured and observed anomalous magnetic moment of the muon~\cite{Pospelov:2008zw} and several intriguing dark matter-related anomalies~\cite{ArkaniHamed:2008qn,Pospelov:2008jd}.  Dark photons can be probed with many types of experiments (e.g.~\cite{Essig:2009nc,Reece:2009un,Strassler:2006im,aubert:2009cp,Batell:2009yf,Baumgart:2009tn,Bjorken:2009mm,Abrahamyan:2011gv,Merkel:2011ze,Essig:2010xa,HPS,Freytsis:2009bh,Wojtsekhowski:2012zq,Batell:2009di,Essig:2010gu,Andreas:2012mt}), including $e^+e^-$ colliders like BaBar, BELLE, future B-factories, and a future linear collider; high-energy proton colliders like the LHC; proton fixed-target experiments that are e.g.~used to produce neutrino beams (like MiniBooNE and MINOS); and electron fixed-target experiments, like those proposed at JLab (including APEX, HPS, and DarkLight), at MAMI, or at VEPP-3.  If dark photons exist in nature, they could provide our only window into a ``hidden sector'' of particles that would otherwise not interact with ordinary matter.  Such hidden-sector particles could include dark matter particles.

\subsubsection*{EIC: e to $\tau$}
With a center of mass energy of up to 160 GeV and a luminosity up to 1.5$\times$10$^{34}$ cm$^{-2}$s$^{-1}$ for electron-proton collisions, the proposed Electron-Ion Collider (EIC) provides a suitable environment for discovering the effects of new physics beyond the SM on electron-proton and electron-ion collisions~\cite{e2tau}.  In particular, a search for $e\rightarrow\tau$ conversion is promising and would complement other searches for lepton flavor violation (LFV) such as $\tau\rightarrow e\gamma,~\mu\rightarrow e\gamma$, and of course $\mu\rightarrow e$ conversion.  Without knowledge of physics beyond the SM, there is \textit{a priori} no reason to expect that stronger limits on LFV in the first and second generations would preclude the possibility of LFV in the first and third generations.  Furthermore, there is a demonstrated possibility for new physics to evade LFV limits from $\tau\rightarrow e\gamma$ and $\mu\rightarrow e\gamma$, satisfy limits from $\mu\rightarrow e$, and still yield $e\rightarrow\tau$ events at a level detectable at the EIC.  The contributions to the $e+p\rightarrow\tau + X$ cross section from photon exchange are small due to present limits from searches for $\tau\rightarrow e\gamma$ and $\tau\rightarrow e\ell\ell$ (where $\ell =e,\mu$)  --  though the $\tau\rightarrow e\ell\ell$ component could receive a large logarithmic enhancement depending on the energy scale of LFV physics.  Contributions from four fermion (two leptons and two quarks) contact operators may be larger as such operators are essentially equivalent to the tree level exchange of a heavy boson.  For the majority of these four fermion operators, the most direct limits were set by searches for leptoquark mediated $e\rightarrow\tau$ at HERA.  Given the high luminosity of the EIC and a reasonable time frame for data collection  --  thus yielding an integrated luminosity of 1000 fb$^{-1}$  --  the EIC could expect to improve on the HERA limits by as much as two orders of magnitude, also surpassing limits on the leptoquark four fermion operators extracted from the upper bound on $\tau\rightarrow e\gamma$.


\section{Standard Model and Beyond: Precision Tests}
Experiments in this category measure quantities that are allowed within the SM, 
establishing values for SM input parameters and/or testing SM predictions at a level of precision that could reveal the presence of NSM dynamics. Of particular interest to the nuclear physics community are the weak decays of the muon and hadrons containing light quarks (pions, neutrons, nuclei), parity-violating asymmetries in the scattering of polarized electrons from unpolarized targets, and the anomalous magnetic moment of the muon.

At the level that they will be pursued in the near future,  the mass reach of these precision probes will overlap or exceed the LHC reach and will therefore be relevant to the construction of the NSM (in this context, even searches with  ``negative results"  will have great value). For example, confirmation of  the current muon g-2 discrepancy~\cite{g-2} with $>\,5\,\sigma$ significance would signal NSM dynamics. If due to heavy particle loop effects, this would indicate that something in the several hundred GeV range  should be seen at the LHC.  

\subsection{Weak Decays of Quarks}
Historically,  weak decays of  quarks inside nuclei  and hadrons  have played a central role in the development of  the Standard Model. Nowadays, in a similar spirit,  precision measurements of charged-current decays of  pions, neutrons, and nuclei provide powerful constraints on many possible realizations of  the New Standard Model, probing scales in the  few TeV range. With increased precision ($< 0.1\%$ level),   in coming years  these  processes might reveal hints of new dynamics and will provide low-energy boundary conditions on new physics scenario that might emerge from  LHC data~\cite{Bhattacharya:2011qm, Bauman:2012fx, Cirigliano:2012ab}.

\subsubsection*{Neutron Decays}
One of the most active areas of neutron nuclear physics concerns the determination of the parameters that describe the $\beta$-decay of the free neutron. Since the process of neutron $\beta$-decay lies at the nexus of a wide range of physics including cosmology, astrophysics, and particle physics (see~\cite{DubbersSchmidt2011} for a recent review), these parameters are of considerable importance. 

Discrepancies among the “best” measurements of the neutron lifetime~\cite{WietfeldtGreene2011, PDG2012} continue to motivate measurements using cold neutron beams at NIST and magnetically stored ultracold neutrons at LANSCE. The most precise cold neutron beam determination of the neutron lifetime was carried out at NIST several years ago.  It was based on the absolute counting of decay protons, with the largest uncertainty attributed to the uncertainty in the calibration of the neutron fluence monitor used to determine the average density in the neutron beam. Recent success in the absolute counting of neutrons at the 0.1\% level has removed the most significant obstacle to improving the precision of this experiment. As a result, a new effort  --  one of five projects identified as high-priority by the 2011 NSAC Neutron Physics subcommittee~\cite{Kumar2011}  --  is underway to measure the neutron lifetime at the 1\,s level using a cold beam at NIST. The apparatus is anticipated to be ready for beam within approximately two years. A next generation neutron lifetime experiment, which could form the start of a path to a lifetime precision of 0.1\,s, is being developed using UCNs from the LANSCE source (the only operating UCN source in the U.S.) .

When the neutron lifetime is combined with measurements of the neutron $\beta$-decay angular correlations (the ``$A$", ``$B$'', or ``$a$'' coefficients), it is possible to extract the weak vector and axial vector coupling constants $g_V$ and $g_A$. These can be compared with nuclear and medium energy data to provide critical checks and constraints on the Standard Model that include tests of the universality of the weak interaction, and to place limits on possible non-Standard Model couplings. 

At LANSCE, the UCNA experiment  --  the first neutron $\beta$-decay correlation measurement to take advantage of the unique properties of ultracold neutrons, and one of five projects identified as high-priority by the 2011 NSAC Neutron Physics subcommittee~\cite{Kumar2011}  --  measures the angular correlation coefficient $A$ of the neutron spin and the electron momentum~\cite{Liu2011, Plaster2012}. The ultimate goal of the experiment is to achieve a total uncertainty of 0.2\% on $A$. Sufficient data to achieve an uncertainty of less than 0.65\% have already been collected and are being analyzed. The intention of the collaboration is to acquire additional data during the 2012 LANSCE accelerator cycle, with systematic improvements that were developed during the 2011 cycle now installed. 

At NIST, the aCORN experiment is aimed at the measurement of the angular correlation between the electron and antineutrino in nuclear $\beta$-decay, $a$. This parameter has the same sensitivity as $A$ to the ratio of the weak coupling constants $g_A$ and $g_V$ and does not require polarized neutrons, but it is less accurately measured. aCORN uses a new method that relies on constructing an asymmetry that directly yields the value of $a$ without requiring precise proton spectroscopy~\cite{Wietfeldt2009}. The goal is to achieve a 2\% relative measurement  on NG-6 before moving the apparatus to the new beam NG-C in early 2013 where the fluence should be roughly five times higher, making a 0.5\% measurement possible. 

The Nab experiment at the SNS  --  one of five projects identified as high-priority by the 2011 NSAC Neutron Physics subcommittee~\cite{Kumar2011}  --   will measure the $a$ and $b$ $\beta$-decay coefficients at the level of 0.1\%~\cite{Pocanic2009}. The apparatus uses a novel electric and magnetic spectrometer~\cite{Bowman2005} to effectively determine the correlation between the electron and neutrino in each decay. This measurement will significantly improve determination of $g_A$ and $g_V$ from neutron decay and place important new limits on tensor and scalar weak couplings. The collaboration has received NSF MRI funding to design and construct the magnetic spectrometer and work is proceeding. A prototype of the custom segmented silicon detectors has been tested using protons with very good results. 

Somewhat more exotic neutron decay processes such radiative neutron decay~\cite{Bernard2004,GardnerHe2012} as well as T-violating (motion-reversal-odd) correlations in ordinary $\beta$-decay~\cite{Drees03} are also studied to provide possible windows on non-Standard Model physics. The emiT collaboration at NIST obtained the best limit ever on T-violation in neutron decay~\cite{Mumm2011}. A new experiment with an improved detection system~\cite{Cooper12} was performed with the aim of measuring the neutron radiative decay spectrum and branching ratio to 1\% accuracy. 

\subsubsection*{Nuclear Decays}

While we expect nuclear $\beta$-decays to be discussed by the FRIB community, there are ongoing experiments at Seattle, Argonne, and elsewhere that are sensitively testing Standard Model $V-A$ expectations.  At Seattle, $\approx 10^9$ $^{6}$He atoms per second can be delivered to experiments where the lifetime has been measured. At Argonne, measurements on the $\beta$-neutrino correlation in the decay of $^8$Li trapped in an ion trap have been completed - the group will soon release the best limit on tensor currents - and a complementary measurement on the mirror decay, that of $^8$B, is in preparation. The combined data set will be sensitive to second class currents and be used to test the CVC hypothesis.

\subsubsection*{Pion Decays}

Two experiments, PEN at PSI~\cite{piref_1a, piref_1b} and PIENU at TRIUMF~\cite{piref_2a, piref_2b}], aim to determine the branching ratio $\pi \to e\nu$, $R^\pi_{e/\mu}$, to a precision of $\sim 5 \times 10^{−4}$, approaching that of the theoretical calculations~\cite{piref_3a, piref_3b, piref_3b,Cirigliano:2007xi} to within a factor of $\sim 6$, as opposed to the current factor of 40.  This tests the fundamental concept of lepton universality, a key assumption in the SM. PEN has completed data taking during this period and PIENU has additional running planned.  First results will be based on current analysis efforts.

Because of large helicity suppression, $R^\pi_{e/\mu}$ is uniquely sensitive to contributions from non-$(V-A)$ physics at large mass scales, $\Lambda_i$.  For example, a measurement of $R^\pi_{e/\mu}$ with just $10^{-3}$ relative precision will probe mass scales of $\Lambda_P \le 1,000$\,TeV and $\Lambda_A \le 20$\,TeV.  Even though $R^\pi_{e/\mu}$ is not directly sensitive to scalar interactions, the latter can induce pseudoscalar terms through loop effects, providing indirect sensitivity at the level $\Lambda_S \le 60$\,TeV, beyond the reach of nuclear $\beta$-decay measurements.  In practice, this sensitivity could be diluted by small Yukawa couplings, or by cancellations of amplitudes; nevertheless precise study of the $\pi \to e\nu$ branching ratio holds significant discovery potential.

\subsection{Neutral Weak Interactions}

The nature of new physics beyond the Standard Model can be revealed at the low-energy precision frontier accessed by JLab, where electroweak studies of parity-violation in electron scattering are planned which will complement the energy-frontier studies at the LHC. In the Standard Model, the vector and axial-vector couplings of fermions to the $Z^0$ boson are precisely predicted as functions of one free parameter: the weak mixing angle $\theta_W$. The effects of new high-energy phenomena can be characterized by new four-fermion contact interaction amplitudes that would contribute to changes in the overall parity-violating interaction of electrons. Measurements of the parity-violating observable $A_{PV}$  --  a polarization-dependent cross-section asymmetry in the scattering of longitudinally polarized electrons from an unpolarized target  --  are sensitive to possible new physics contributions to multi-TeV mass scales~\cite{PVES1}.  Recent measurements of $A_{PV}$ have focused on exposing details of nucleon or nuclear structure using the weak neutral current ($Z^0$), but in combination have been found to provide an important constraint on the weak vector charge of the first generation quarks~\cite{PVES2}.  This constraint will be further improved when results are available from the recently completed QWeak experiment, measuring the proton weak charge to a precision of 4\% with a corresponding sensitivity to possible new contact interactions to a scale (energy scale over coupling) of  $\Lambda/g \sim 2.5$\,TeV~\cite{PVES3}. 

The 12 GeV upgrade at JLab provides a unique opportunity to measure $A_{PV}$ in M{\o}ller scattering with a precision approaching 2\%, a factor of 5 improvement over previous results~\cite{PVES4}. M{\o}ller scattering is a purely leptonic process and a particularly clean way to test the Standard Model with high precision. The MOLLER experiment has been proposed to have a reach in sensitivity for new four-electron contact interaction amplitudes as small as $1.5\times 10^{−3}\times G_F$. This corresponds to a sensitivity of $\Lambda/g = 7.5$\,TeV, providing the most sensitive probe of new flavor and CP-conserving neutral current interactions in the leptonic sector until the advent of a linear collider or a neutrino factory. If  LHC finds evidence for physics beyond the Standard Model, it would become particularly important to test the agreement between the directly measured value of Higgs mass $m_H$, and that inferred from the measurements of fundamental electroweak parameters $\sin^2 \theta_W$, the mass of the $W$ boson, and the mass of the top quark. With precision comparable to the best high-energy measurements of $\sin^2 \theta_W$, MOLLER would be the first low-$Q^2$ measurement capable of providing a meaningful $m_H$ constraint through loop-level corrections to the scattering process.

The SOLID experiment has also been proposed at JLab to study parity-violating deep inelastic scattering with a new solenoidal magnetic spectrometer system~\cite{PVES5}. The use of a deuterium target minimizes uncertainties due to quark distribution functions making the measurement sensitive to weak axial charge of the first generation quarks. The  SOLID experiment would improve the determination of these couplings by a factor of 6 to 8 over present knowledge, with complementary sensitivity to new TeV-scale physics. The optimum strategy would achieve simultaneous $A_{PV}$ measurements in narrow kinematic $(x_B, Q^2)$ bins to provide precision measurements of the quark weak couplings while simultaneously providing new insights into nucleon structure in the valence region.

\subsection{Lepton Properties and Interactions}

Significant progress has been made in a number of experiments within the U.S. precision muon physics program since the last LRP.

\paragraph{TWIST} The TWIST experiment at TRIUMF studied the detailed Michel parameters from positive muon decay at rest in order to explore possible extensions to the Standard Model, in particular, right-handed muon terms in the weak interaction. The experiment completed data taking and analysis and published results on the parameters $\rho, \delta$ and $P_{\mu}\xi$, achieving on average an order of magnitude improvement in precision compared to previous work~\cite{Bueno11, Bayes11}. Taken together with a global analysis of muon decay parameters, the results -- which agree with the SM -- provide significant new limits on right-handed currents and support the $V-A$ decay description.

\paragraph{MuLan} The MuLan experiment at PSI completed data taking and analysis of more than $2 \times 10^{12}$ positive muon decays to determine the lifetime.  Two runs, using vastly different target strategies, were in agreement to a precision of 1.3\,ps.  Combined, the lifetime has been determined to an unprecedented 1\,ppm relative precision (2.2\,ps), the most precise lifetime ever measured~\cite{Webber11}.  It is used to determine the Fermi Constant, $G_F$, which together with updated theoretical calculations, is now known to 0.5\,ppm.  The positive muon lifetime is also used as the benchmark lifetime in the muon capture program, described next.

\paragraph{MuCap} The MuCap experiment measured the singlet $\mu-p$ capture rate, $\Lambda_S$ in an ultra-pure hydrogen time projection chamber (TPC).  The choice of a gas target  --  absent both isotopic impurities and other impurities to unprecedented levels  --  allows for an unambiguous measurement of the capture rate in the atomic state, rather than the molecular state, which plagued previous muon capture experiments.  The capture rate is deduced from the difference between the lifetime measured by MuCap~\cite{mucap1,mucap2} and the unaffected positive muon lifetime measured by MuLan~\cite{Webber11}. The capture rate is predicted precisely by QCD-based low-energy effective theory, which established the weak-nucleon pseudoscalar coupling, $g_P$.  The new experimental result, together with updated radiative corrections also published  during this period~\cite{MuCAP_THEORY}, agrees exactly with the theory, resolving a long standing puzzle.

\paragraph{MuSun} The MuSun experiment was proposed and mounted during this period~\cite{MUSUN}.  It is similar to MuCap, except that it will measure the capture rate in cold deuterium gas, using a new cryo-TPC.  Several fundamental astrophysics processes depend on the weak interaction in deuterium, for example $pp$ fusion in the sun, and the neutrino breakup reactions as measured in the SNO detector. These tiny cross sections rely on theory and they share a common effective theory approach with a common low-energy constant (LEC), for example, the poorly determined LEC $L1A$.  MuSun -- already taking physics data -- will determine this LEC, which can then be used to refine the predictions in these other fundamental reactions.

\paragraph{g-2} The new $g-2$ experiment at Fermilab (E989)~\cite{g-2_redux} is derived from the very successful BNL E821 experiment, which determined the muon anomaly to a statistics-limited precision of 0.54\,ppm. E989  is designed to acquire approximately 20 times the earlier data set in a little over a year of running.  The polarized $3.1\,$GeV/$c$ muon beam is provided by the existing Fermilab accelerator complex, using extra proton batches in the supercycle that feeds the NO$\nu$A experiment; that is, it will run parasitic to the neutrino program.  A high-purity, background-free muon beam will be directed to the relocated BNL storage ring, which will be housed in a new temperature-controlled building at the ``Muon Campus'' that anchors a major part of the new Intensity Frontier initiative at the lab.  One of the major challenges is moving the ring; that task has begun in earnest in 2012 with expected shipping of the delicate superconducting coils in 2013. Nuclear physicists play key roles in this experiment and many groups involved will provide the detectors, electronics and data acquisition systems required for a next-generation experiment. With a precision goal of 0.14\,ppm, the impact -- assuming the same central value as BNL found -- would exceed the $5\,\sigma$ discovery threshold.  At present many groups are involved in R\&D efforts and design work for the beams and muon storage dynamics.

\subsection{Weak Interactions Between Quarks}

The last decade has seen advances in the quantitative understanding of nuclei, especially few body systems, and in the connection between nuclear physics and quantum chromodynamics (QCD). Low energy properties of nucleons and nuclei, such as weak interactions in $n-A$ systems, low-energy $n-A$ scattering amplitudes, and the internal electromagnetic structure of the neutron are becoming calculable in the SM despite the strongly interacting nature of these systems. These theoretical developments are motivating renewed experimental activity to measure undetermined low-energy properties, such as the weak interaction amplitudes between nucleons, and to improve the precision of other low-energy neutron measurements. Independent of the theoretical model, several experimental approaches are required to narrow the range of the predictions. 

The NPDGamma experiment  --  one of five projects identified as high-priority by the 2011 NSAC Neutron Physics subcommittee~\cite{Kumar2011}  --   measures the gamma-ray asymmetry $A_{\gamma}$ in the process $p(n,\gamma)d$~\cite{Gericke2011, Bowman2007}. The experiment had a successful run at LANL and an upgraded version is now in full operation at the SNS. The SNS neutron flux provides a factor of $\approx 100$ increase in data rate and a projected $A_{\gamma}$ sensitivity of $10^{-8}$. 

NPDGamma will provide one constraint to the leading order low energy constants needed to verify that we understand the manifestation of parity violation in hadrons.  A complementary measurement, providing a different constraint, comes from the circularly polarized breakup of the deuteron.
This may be possible with an intensity upgrade to the High Intensity Gamma Source at the Triangle Universities Nuclear Laboratory (HIGS2).

It is anticipated that between the NPDgamma experiment and the Nab experiment (discussed in Section~4.1), there will be time to perform a measurement of the fragmentation asymmetry in the process $^3{\rm He} (n,p) ^3\!{\rm H}$ at the ORNL FNPB. This experiment will use much of the infrastructure currently in place for NPDGamma. 

A statistics limited bound on the parity violating neutron spin rotation in liquid helium was recently obtained~\cite{Snow2011}. The experiment ran on the NG-6 beam at NIST where it was statistics limited~\cite{Snow2011}. It is currently undergoing an upgrade and will run on the new higher-flux NG-C beam. 

\subsection{Weak Probes of Nuclear Physics and Astrophysics}

Novel studies of lepton-hadron weak interactions using parity-violating electron scattering and neutrino-nucleus scattering, as well as sensitive detection of terrestrial and astrophysical sources of neutrinos are leading to novel insights into the structure of dense nuclear matter, with important implications for nuclear physics and astrophysics.

Analogous to the iconic use of electron scattering to measure charge densities of nuclei, it has now become possible to use parity-violating electron scattering to measure the weak densities of heavy nuclei. Because the $Z$ boson couples to neutrons and has a very small coupling to protons, the weak density is a very clean measure of the neutron density. The PREX experiment at JLab has made the first such electroweak measurement of $^{208}$Pb~\cite{Abrahamyan:2012gp} and a subsequent measurement scheduled for 2015 will measure the neutron RMS radius to a fractional accuracy of 1\%. A new measurement on $^{48}$Ca is also being planned. These measurements are a critical input to many-body nuclear theory and an important constraint on the density dependence of the symmetry energy, which has broad applications in nuclear physics and astrophysics, including providing input to estimates of neutron star properties~\cite{Tsang:2012se}.

Their tiny masses and incredibly weak interaction strengths make neutrinos our ``hidden messenger'', allowing experiments designed to measure neutrino properties to also probe otherwise unreachable regions: from the depths of the Earth beneath our feet to the very core of the Sun, and even into far-distant supernovae. Surprisingly, these particles, so far removed from everyday experience, may also prove to useful monitors for compliance with nuclear non-proliferation treaties.

U.S. nuclear physics has pioneered the use of neutrinos as probes of the Sun, having contributed heavily to the SNO experiment, which resolved the Solar Neutrino Problem in 2001 and has since produced high-precision measurements of the $^8$B solar neutrino flux. This opens the door to other precision measurements, including testing the luminosity constraint, resolving the solar metallicity, and either confirming the details of MSW or potentially discovering new physics in the vacuum-matter transition region.  Borexino (mainly NSF supported) has produced many exciting results, including a precision measurement of the $^7$Be solar neutrino flux~\cite{Bellini2011} and the first hint at a direct observation of PEP neutrinos~\cite{Bellini2012}, as well as the tightest bounds on the CNO flux.  SNO+ is expected to start taking data in the next couple of years and, with its larger size and deeper location, should be able to address many remaining questions.  Future proposed experiments include LENS, which proposes to use charged-current interactions on indium, thus allowing the extraction of an energy spectrum for low energy (primarily $pp$) solar neutrinos, and large liquid noble gas dark matter experiments such as CLEAN, which could also achieve percent-level precision on the $pp$ flux using elastic scattering. 

KamLAND and Borexino have detected neutrinos originating within the earth (``geo neutrinos")~\cite{geo_neutrinos}.  Both will continue to take data and improve the precision on these measurements, and when SNO+ turns on a third geographical location should be added, which will improve constraints on models of radiogenic heat production in the Earth.

Core-collapse supernovae produce a vast flux of neutrinos of all flavors with energies in the few tens-of-MeV range over a few tens of seconds.  The detection of this intense neutrino burst for a nearby supernova would bring tremendous information about neutrino physics, about the astrophysics of core collapse, and about supernova nucleosynthesis~\cite{ScholSN}.  Detectors with diverse flavor sensitivities are highly desirable. U.S. nuclear physicists are involved in several experiments sensitive to supernova neutrinos (such as KamLAND, IceCube, Borexino, SNO+), as well as dedicated supernova experiments such as HALO. 

As mentioned in Section~3.2, the ORNL Spallation Neutron Source is also a potent neutrino source. As such, it is ideal for measurements of $\nu$-nucleus cross sections in the few tens-of-MeV range in a variety of targets which is vital for understanding core-collapse dynamics and supernova nucleosynthesis, which are highly sensitive to $\nu$ processes. Such measurements would also enhance our ability to extract information about $\nu$ mixing properties (in particular, the mass hierarchy) from the observation of a galactic supernova. Another interesting possibility is the detection of nuclear recoils from coherent elastic $\nu$-nucleus scattering, which is within the reach of the current generation of low-threshold detectors~\cite{Scholberg:2005qs}. This reaction is also important for supernova processes and detection and has excellent prospects for Standard Model tests; even a first-generation experiment has sensitivity beyond the current best limits on non-Standard Model interactions of neutrinos and quarks~\cite{Scholberg:2009ha}.

MINER$\nu$A is an on-axis neutrino-nucleus scattering experiment at Fermilab's NuMI (Neutrinos at the Main Injector) beamline that will measure interaction cross-sections and event kinematics in exclusive and inclusive states to high precision. Although an HEP-funded experiment, there are several (NSF) NP-funded collaborators in the U.S. This experiment will also examine nuclear effects and parton distribution functions using a variety of target materials. This information is crucial to a quantitative understanding of neutrino production, backgrounds and calibration.

The reactor neutrino oscillation measurements described in Section~3.2 are closely connected to an important goal for international nonproliferation and nuclear security. For these experiments, the underlying theory and technology provide a new tool for monitoring the inventory of fissile material being consumed and generated in nuclear reactors~\cite{reactor_nu_monitors}.  In particular, the reactor antineutrino flux and spectrum are correlated with the changing inventories of fissile materials in the cores of nuclear reactors. Monitoring the flows of these weapons-usable materials in the civilian nuclear fuel cycle is a priority for the International Atomic Energy Agency (IAEA).  Reactor antineutrino flux predictions and  experiments must take into account the systematic effect of the changing fissile inventory, as large as 10-15\% over the course of the fuel cycle. This theoretical framework, and the experimental outputs from detectors, can be used to solve the inverse problem of estimating inventories when provided with a measured spectrum, or of simply confirming the operational status and thermal power of the reactor in near-real-time.  Such information is not otherwise easily accessible to the IAEA, and as a result considerable interest in the concept of antineutrino monitoring has developed in the physics and nonproliferation communities.  There are now roughly one dozen small scale applied antineutrino technology and theoretical efforts active worldwide, all closely connected to the fundamental physics community, and the IAEA has recently convened a working group to consider the possible application of the technology to its own reactor monitoring regime. The opportunity to provide a new capability to the international community, that fosters the peaceful use of atomic energy, is an important benefit of ongoing work on reactor-based neutrino oscillation experiments.


\clearpage
\section{Facilities for Fundamental Symmetries and Neutrinos}\label{sec:facilities}

The U.S.\ nuclear physics community employs a large number of
different facilities for research in fundamental symmetries and
neutrinos.  Tables
\ref{tab:neutron_facilities} -- \ref{tab:reactors_very_short_baseline} below
summarize the current and projected future status of: (a) accelerator-
and reactor-based facilities for neutron physics; (b) underground labs
for neutrino physics; (c) accelerator-based facilities for muon
physics; (d) a general survey of facilities for other types of
research in nuclear $\beta$-decay, storage ring EDMs, kaon physics,
etc.; and (e) U.S.\ research reactor facilities under consideration
for a very short baseline reactor neutrino experiment.  The content of
these tables is not comprehensive, and is
primarily focused on those facilities utilized by the U.S.\ nuclear
physics community. However, comparison with a number of international
facilities is also provided for context.

\begin{table}[h!]
\begin{center}
\caption[Summary of accelerator- and reactor-based facilites for neutron
physics research.]{Summary of accelerator- and reactor-based facilites for neutron
physics research.  `CN' denotes `cold neutrons', `UCN' denotes
`ultracold neutrons'.}
\begin{tabular}{lll} \hline\hline
\textbf{Facility (Funding)}& \textbf{Type}& \textbf{Status and Assumptions} \\ \hline
\multicolumn{3}{c}{\textbf{ --  U.S.-Based Facilities  -- }} \\ \hline
NIST: NCNR& 20 MW reactor, & Operating,
  high-flux beam available by end of 2012, \\
(Dept.\ of Commerce)& CN beams& neutron $\beta$-decay,
  hadronic PV, crystal nEDM \\ \hline
SNS: FNPB& Spallation source,& Operating, hadronic parity violation,
  neutron \\
(BES)& CN beams& $\beta$-decay, nEDM Experiment (superfluid
  $^4$He UCN) \\ \hline
LANL: LANSCE& Spallation-driven, solid& Operating, only source
  of extracted UCN in U.S. at \\
(DOE NP, LDRD)& deuterium UCN source& present, neutron $\beta$-decay, nEDM
  Experiment R\&D \\ \hline
NCSU: PULSTAR& 1 MW reactor, solid& Under development, UCN production
  by late-2012, \\
(NSF, DOE NP)& deuterium UCN source& nEDM Experiment R\&D, small-scale
  experiment \\ \hline
\multicolumn{3}{c}{\textbf{ --  International Facilities  -- }} \\ \hline
PSI& Spallation-driven, solid& Under development, operating by late-2012, \\
& deuterium UCN source& PSI-based nEDM Experiment \\ \hline
ILL& 58 MW reactor, CN& Operating, cryoEDM Experiment, $\beta$-decay,
  gravity, \\
& beams, UCN sources& high-density UCN source under development \\ \hline
Munich FRM-II& 20 MW reactor, CN& Solid deuterium UCN source operational
  by late-  \\
& beam, UCN source& 2012, MEPHISTO CN beam for PERC $\beta$-decay \\ \hline
RCNP/TRIUMF& Spallation-driven,& Under development at RCNP, move to\\
& superfluid $^4$He UCN& TRIUMF by 2015, TRIUMF nEDM Experiment \\ \hline
J-PARC& Spallation-driven, solid& Under development, J-PARC nEDM
  Experiment \\
& deuterium UCN source \\ \hline
PNPI& 15 MW reactor,& Under development, neutron $\beta$-decay, crystal \\
& superfluid $^4$He UCN& nEDM, long-range forces \\ \hline\hline
\end{tabular}
\label{tab:neutron_facilities}
\end{center}
\end{table}

\begin{landscape}


\begin{table}[h!]
\begin{center}
\caption[Summary of underground labs for neutrino physics research.]{Summary of underground labs for neutrino physics research.  In the Experiments column, `DM' denotes `Dark Matter',  `(P)' denotes `Proposal'.
[See National Research Council Report on ``An Assessment of the Deep Underground Science and Engineering Laboratory'' \href{http://www.nap.edu/openbook.php?record_id=13204&page=1}{\tt {http://www.nap.edu/openbook.php?record\_id=13204\&page=1}} for more details.]}
\begin{tabular}{lcclcc} \hline\hline
\textbf{Underground Lab}& \textbf{Depth}& \textbf{Rock Type}& \textbf{U.S.-Funded Experiments}& \textbf{Volume}& \textbf{Area} \\
~~\textbf{Location, Purpose, Funding}& \textbf{[mwe]}& \textbf{U/Th [ppm]}& \textbf{and Proposals}& \textbf{[m$^3$]}& \textbf{[m$^2$]}  \\ \hline
\multicolumn{6}{c}{\textbf{ --  U.S.-Based Underground Labs  -- }} \\ \hline
Sanford Underground Research Facility& Davis Campus:& Medamorphic& LUX DM (2011 -- 2015),& 6395& 2731 \\
$[$Levels 0 -- 8000 every 150 feet$]$& 4300& 0.2/0.3& Majorana $0\nu\beta\beta$ (2011 -- 2018), \\
~~South Dakota, Research,& & & CUBED Low Background (2012 -- ), \\
~~DOE HP and NP, LBNL& & & DIANA (P), LZ DM (P), LBNE (P) \\ \hline
Waste Isolation Pilot Plant& 1600& Salt Dome& EXO-200 $0\nu\beta\beta$ (2011 -- 2016+),& 5050& 920 \\
~~New Mexico, Waste Storage& & 0.1/0.1& DMTPC DM (2012 -- 2017+), \\
~~DOE&&& Low Background Counting, R\&D \\ \hline
Soudan& 2090& Ely Greenstone& MINOS+ (2012 -- 2015+),& 25230& 2070 \\
~~Minnesota, Research,& & 0.2/0.9& Super-CDMS DM (2012 -- 2016+),& \\
~~DOE HEP, U.\ Minnesota& & & CoGeNT DM (2012 -- 2016+),& \\
&&& Low Background Counting \\ \hline
Kimballton Underground Research Facility& 1450& Limestone& mini-LENS, Low Background,& & 325 \\
~~Virginia, Active Mine,& & & MALBEK, DIANA (P),& \\
~~Virginia Tech& & & $\beta\beta$ Excited States, Depleted Ar& \\ \hline
Oroville& 300& Earthen Dam,& Low Background Counting& 1500& 150 \\
~~California, Hydro-Power, LBNL& & Water, 5/10& \\ \hline
Proposed: Cascade Tunnel& 1430& Granite& \\
~~Washington State, Former Railway Tunnel \\ \hline
\multicolumn{6}{c}{\textbf{ --  International Underground Labs with U.S.\ Involvement in Experiments  -- }} \\ \hline
SNOLAB& 6000& Granite and& SNO+ ($0\nu\beta\beta$, solar neutrinos,& 29555& 3055 \\
~~Ontario, Canada, Ni Mine,& & Ni Sulfites& geoneutrinos, supernova, etc.),& \\
~~CFI, NSERC& & 5/10& broad DM program, Supernova \\ \hline
LGNS Gran Sasso& 3700& Diverse& Borexino, CUORE,& 180000& 18000 \\
~~Abruzzi, Italy, Highway Tunnel,& & 2/7& Xenon-100, 1-Ton (P) \\
~~LGNS \\ \hline
Kamioka& 2700& Carbonate,& T2K, Super-K,& 56500& 1527 \\
~~Japan, Research,& & Igneous, 5/10& KamLAND-Zen $0\nu\beta\beta$\\
~~Japanese Funding Agencies, ICRR \\ \hline\hline
\end{tabular}
\label{tab:underground_labs}
\end{center}
\end{table}

\end{landscape}

\begin{table}[h!]
\begin{center}
\caption{Summary of accelerator-based facilities for muon physics research.}
\begin{tabular}{ll} \hline\hline
\textbf{Facility}& \textbf{Status and Assumptions} \\ \hline
\multicolumn{2}{c}{\textbf{ --  U.S.-Based Facilities  -- }} \\ \hline
FNAL Muon Campus& Upgrades for Booster to supply 8 GeV proton batches at
  15 Hz, protons to \\
(DOE HEP)& Recycler (not Main Injector), split into bunches (via upgraded
  r.f.) \\
& ~~~For $g-2$, bunches extracted to antiproton target for pion/muon
  production, \\
& ~~~~~~muons transported to Debuncher (renamed Delivery Ring), then \\
& ~~~~~~extracted to $g-2$ ring \\
& ~~~For Mu2e, proton bunches stored in Delivery Ring, resonantly
  extracted \\
& ~~~~~~to Mu2e Production Target \\
& Key Assumptions: \\
& ~~~(1) Ground-breaking on Muon Campus commences Q1 FY12 \\
& ~~~(2) Muon Campus completed by FY16/17 \\
& ~~~(3) Seeking nuclear physics MRI support for $g-2$
  detector/DAQ/electronics \\
& ~~~(4) Continued nuclear physics support for groups on
  $g-2$ and Mu2e \\ \hline
\multicolumn{2}{c}{\textbf{ --  International Facilities  -- }} \\ \hline
PSI& 590 MeV (1.4 MW) proton ring cyclotron, pion/muon production on \\
& graphite targets, significant U.S.\ involvement in MuLan, MuCap, \\
& MuSun (running), U.S.-led proposal for $\mu p$ elastic scattering (proton \\
& charge radius); also MEG (running), PSI proposal for
  $\mu \rightarrow eee$ \\ \hline
J-PARC& 3 GeV (1 MW) proton beam for muon production on graphite targets, \\
& COMET experiment (similar to U.S.\ Mu2e) \\ \hline\hline
\end{tabular}
\label{tab:muon_facilities}
\end{center}
\end{table}

\begin{table}[h!]
\begin{center}
\caption{Summary of other facilities for fundamental symmetries research.}
\begin{tabular}{ll} \hline\hline
\textbf{Facility}& \textbf{Status and Assumptions} \\ \hline
\multicolumn{2}{c}{\textbf{ --  Nuclear $\beta$-Decay  -- }} \\ \hline
Texas A\&M Cyclotron& TREX upgrade to cyclotron, new TAMUTRAP Penning
  trap for \\
& new superallowed $p$ emitter $ft$ values \\ \hline
U.\ Washington& Tandem Van de Graaff for $^6$He production, $\beta$-decay
  studies \\ \hline
Argonne& $\beta$-decay Paul trap coupled to ATLAS, $^8$Li and $^8$B
  $\beta$-decay \\ \hline
NSCL/FRIB& BECOLA endstation, polarized nuclei $\beta$-decay of
  $^{21}$Na, $^{23}$Mg, $^{36}$K, $^{37}$K \\ \hline
LBNL& Cyclotron for $^{21}$Na $\beta$-decay studies \\ \hline
TRIUMF& TRINAT MOT for $^{37}$K and $^{38\mathrm{m}}$K $\beta$-decay,
  Fr trapping for anapole moment \\ \hline
\multicolumn{2}{c}{\textbf{ --  Other  -- }} \\ \hline
JLab & After 12 GeV Upgrade: 11 GeV MOLLER $\sin^2 \theta_W$, PVDIS SOLID $C_{iq}$'s \\ \hline
COSY& Joint BNL proton/deuteron EDM R\&D effort with COSY Deuteron \\
& EDM Collaboration \\ \hline
J-PARC Kaon Physics& 30 GeV proton beam for kaon production, U.S.-led
  TREK experiment \\ \hline
FNAL Project X& Proposed; Stage 1: New superconducting CW 1 GeV proton
  linac; \\
& neutrino physics, kaon physics, proton EDM, CN/UCN sources \\
& for $n$-$\overline{n}$ oscillations, nEDM, etc.; construction later
  this decade; \\
& operations early next decade \\ \hline
SNS Neutrino Facility& Proposed; neutrinos from pion decay at rest and
  muon decay, \\
& OscSNS and $\nu$-nucleus scattering \\ \hline\hline
\end{tabular}
\label{tab:other_facilities}
\end{center}
\end{table}

\begin{table}[h!]
\begin{center}
\caption[Power, accessible baselines, fuel cycles, and
reactor-off times at U.S.\ reactor facilities.]{Reactor powers, accessible baselines, fuel cycles, and
reactor-off times at various U.S.\ reactor facilities. The down-time
includes estimates for seasonal shutdowns and maintenance periods.
U.S.\ research reactors provide some of the highest $\overline{\nu}_e$
flux worldwide with the shortest baselines.}
\begin{tabular}{lllllll}\hline\hline
\textbf{Reactor}& \textbf{Power}& \textbf{Baselines}& \textbf{Cycle On}&
  \textbf{Cycle Off}& \textbf{Down-Time}& \textbf{Ref.} \\
& \textbf{[MW$_\mathrm{th}$]}& \textbf{[m]}& \textbf{[Days]}&
  \textbf{[Days]}& \\ \hline
NIST& 20& 4 -- 13& 42& 10& $\sim 32$\%& (a) \\ \hline
HFIR& 85& 6 -- 8& 24& 18& $\sim 50$\%& (b) \\ \hline
ATR& 250 (licensed)& 7 -- 8 (restricted)& 48 -- 56& 14 -- 21& $\sim 27$\%& (c) \\
& 110 (operational)& 12 -- 20 (full access)& \\ \hline\hline
\multicolumn{7}{l}{(a) 2012 Operations Schedule, NIST Center
for Neutron Research,} \\
\multicolumn{7}{l}{~~~~~\url{http://www.ncnr.nist.gov/ns_schedule.html},
(2012).} \\
\multicolumn{7}{l}{(b) HFIR Operating Forecast \& Planning
Schedule, Oak Ridge National Laboratory,} \\
\multicolumn{7}{l}{~~~~~\url{http://neutrons.ornl.gov/facilities/HFIR/pdf/rolling15month.pdf}, (2012).} \\
\multicolumn{7}{l}{(c) FY 2009 Advanced Test Reactor National Scientific User Facility Users' Guide, ATR National} \\
\multicolumn{7}{l}{~~~~~Scientific Users Facility,} \\
\multicolumn{7}{l}{~~~~~\url{https://secure.inl.gov/atrproposal/documents/ATRUsersGuide.pdf}, (2009).}
\end{tabular}
\label{tab:reactors_very_short_baseline}
\end{center}
\end{table}


\clearpage
\section{International Context}

Scientific progress in fundamental symmetries and neutrinos, much like the rest of scientific endeavor, often requires the combined efforts of many countries. International cooperation and collaborations in science have the added benefit of being precursors to international connections in social areas and commerce. 

International collaboration in neutrino physics is particularly important as major underground laboratories that provide low background from cosmic rays and hence provide sensitivity to rare processes are mostly located outside the United States. Major investments in new or upgraded facilities that explore neutrinos and fundamental symmetries have been made in China, Europe,  and Japan over the last decade. U.S. scientists continue to take advantage of these facilities in a very beneficial way. But if the U.S. is to remain a global leader in this subfield it must simultaneously develop its own facilities while globally pursuing scientific opportunities. 

Neutrino experiments based outside the U.S. in which U.S. nuclear physicists play a leading role include the CUORE neutrinoless double $\beta$-decay experiment in Gran Sasso, the KamLAND-Zen  neutrinoless double $\beta$-decay experiment in Kamioka, and the Daya Bay reactor neutrino experiment at  Guangdong Province in China. In addition, U.S. physicists are involved in most of the experiments at SNOLAB and play a leading role in many of them. The KATRIN collaboration, which has a strong U.S. involvement, unites in Germany the world-wide expertise in tritium $\beta$-decay to probe the absolute mass scale of neutrinos. 
The {\sc MAJORANA} Collaboration is working cooperatively with the GERDA Collaboration in Europe towards an international ton-scale germanium double $\beta$-decay experiment that combines best features of the two programs. 

Canadian research in subatomic physics is centered at the TRIUMF laboratory. TRIUMF's flagship program is in rare isotope beams. In addition to research in nuclear structure and reactions, current and planned rare-isotope beam facilities at TRIUMF would allow tests of fundamental symmetries, such as atomic electric dipole moments enhanced by octupole nuclear shapes. U.S. researchers are involved in the radon EDM experiment at TRIUMF. In addition, U.S. physicists participate at the nuclear $\beta$-decay program at TRIUMF.  On the European side, there is a significant U.S. involvement at the Paul Scherrer Institute, PSI,  Switzerland, primarily centered on muon physics. 

International collaboration in theoretical nuclear physics research is also commonplace. The European Centre for Nuclear Physics and Related Areas (ECT*) at Trento, Italy, aims to facilitate in-depth research on topical problems at the forefront of contemporary developments in nuclear physics and related areas by arranging workshops and collaboration meetings. It also contributes to the training of next generation of nuclear scientists through several programs aimed at the graduate students. All these programs have a good participation from the U.S., and the Scientific Advisory Board of ECT* always includes a U.S.-based nuclear physicist. Currently the U.S. representative is the Chair of this Board. Programs focused on collaborations with individual countries in low-energy theoretical research, such as JUSTIPEN with Japan and FUSTIPEN with France, also indirectly help collaborative efforts in neutrinos and fundamental symmetries. 

The global nuclear science community also positively impacts nuclear physics research activities within the U.S. For example, international participation at INT has been crucial to its success. Additionally, many U.S.-based experiments are also international collaborations with active participation of scientists from the broader international community. Examples include EXO, {\sc Majorana}, and many of the JLab experiments among others.


\clearpage
\appendix
\section*{Links to Write-ups Submitted to The Fundamental Symmetries and Neutrinos Workshop, August 10-11, 2012}
\begin{itemize}

\item ``{\it Opportunities for Neutrino Physics at the Spallation Neutron Source}'' \\ \hspace*{0.5 in}
	 \href{http://www.phy.ornl.gov/funsym/positions/sns_neutrinos.pdf}{\tt http://www.phy.ornl.gov/funsym/positions/sns\_neutrinos.pdf}

\item ``{\it OscSNS: A Precision Neutrino Oscillation Experiment at the SNS}'' \\ \hspace*{0.5 in}
	 \href{http://www.phy.ornl.gov/funsym/positions/oscsns.pdf}{\tt http://www.phy.ornl.gov/funsym/positions/oscsns.pdf}

\item ``{\it Opportunities to Advance Fundamental Symmetries Research with Project X}'' \\ \hspace*{0.5 in}
	 \href{http://www.phy.ornl.gov/funsym/positions/ProjectXFunSymNeutrinos.pdf}{\tt http://www.phy.ornl.gov/funsym/positions/ProjectXFunSymNeutrinos.pdf}

\item ``{\it Search for a Neutron Electric Dipole Moment at the SNS}'' \\ \hspace*{0.5 in}
	 \href{http://www.phy.ornl.gov/funsym/positions/nEDM-position-paper.pdf}{\tt http://www.phy.ornl.gov/funsym/positions/nEDM-position-paper.pdf}

\item ``{\it Searches for Electric Dipole Moments}'' \\ \hspace*{0.5 in}
	 \href{http://www.phy.ornl.gov/funsym/positions/FundSymm-EDM.pdf}{\tt http://www.phy.ornl.gov/funsym/positions/FundSymm-EDM.pdf}

\item ``{\it Opportunities for the Precision Study of Reactor Antineutrinos at Very Short Baselines at U.S. Research Reactors}'' \\ \hspace*{0.5 in}
	 \href{http://www.phy.ornl.gov/funsym/positions/NSAC_sterile_whitepaper_v2.pdf}{\tt http://www.phy.ornl.gov/funsym/positions/NSAC\_sterile\_whitepaper\_v2.pdf}

\item ``{\it Fundamental Physics with Cold Neutrons}'' \\ \hspace*{0.5 in}
	 \href{http://www.phy.ornl.gov/funsym/positions/NSAC2012_ColdNeutronsv3.pdf}{\tt http://www.phy.ornl.gov/funsym/positions/NSAC2012\_ColdNeutronsv3.pdf}

\item ``{\it Neutrinos from STORed Muons: $\nu$STORM}'' \\ \hspace*{0.5 in}
	 \href{http://www.phy.ornl.gov/funsym/positions/nuSTORM_1a.pdf}{\tt http://www.phy.ornl.gov/funsym/positions/nuSTORM\_1a.pdf}

\item ``{\it The Mu2e Experiment at Fermilab: a Search for Charged Lepton Flavor Violation}'' \\ \hspace*{0.5 in}
	 \href{http://www.phy.ornl.gov/funsym/positions/Mu2eforNSAC.pdf}{\tt http://www.phy.ornl.gov/funsym/positions/Mu2eforNSAC.pdf}

\item ``{\it The Sanford Underground Research Facility at Homestake}'' \\ \hspace*{0.5 in}
	 \href{http://www.phy.ornl.gov/funsym/positions/SURF_whitepaper_DNP_short.pdf}{\tt http://www.phy.ornl.gov/funsym/positions/SURF\_whitepaper\_DNP\_short.pdf}

\item ``{\it Parity Violation in Photonuclear Reactions at HIGS}'' \\ \hspace*{0.5 in}
	 \href{http://www.phy.ornl.gov/funsym/positions/parityviolationatHIGS.pdf}{\tt http://www.phy.ornl.gov/funsym/positions/parityviolationatHIGS.pdf}

\item ``{\it Symmetry Violation in Nuclei}'' \\ \hspace*{0.5 in}
	 \href{http://www.phy.ornl.gov/funsym/positions/Gudkov-SymmetryViolationNuclei.pdf}{\tt http://www.phy.ornl.gov/funsym/positions/Gudkov-SymmetryViolationNuclei.pdf}

\item ``{\it Neutrino Physics with SNO+}'' \\ \hspace*{0.5 in}
	 \href{http://www.phy.ornl.gov/funsym/positions/SNO+FunSymmDNP.pdf}{\tt http://www.phy.ornl.gov/funsym/positions/SNO+FunSymmDNP.pdf}

\item ``{\it Search for Neutrinoless Double $\beta$-Decay with CUORE}'' \\ \hspace*{0.5 in}
	 \href{http://www.phy.ornl.gov/funsym/positions/Kolomensky-nsac_funsym2012_v2.pdf}{\tt http://www.phy.ornl.gov/funsym/positions/Kolomensky-nsac\_funsym2012\_v2.pdf}

\item ``{\it  Search for Neutrinoless Double $\beta$-Decay of Germanium-76}'' \\ \hspace*{0.5 in}
	 \href{http://www.phy.ornl.gov/funsym/positions/MAJORANA_FS_Neu_2012.pdf}{\tt http://www.phy.ornl.gov/funsym/positions/MAJORANA\_FS\_Neu\_2012.pdf}

\item ``{\it  Next Generation Neutrinoless Double $\beta$-Decay Experiments}'' \\ \hspace*{0.5 in}
	 \href{http://www.phy.ornl.gov/funsym/positions/DoubleBetaDecay_v7.pdf}{\tt http://www.phy.ornl.gov/funsym/positions/DoubleBetaDecay\_v7.pdf}

\item ``{\it  Storage Ring Proton EDM}'' \\ \hspace*{0.5 in}
	 \href{http://www.phy.ornl.gov/funsym/positions/YkS_proton_EDM1-standalone.pdf
}{\tt http://www.phy.ornl.gov/funsym/positions/YkS\_proton\_EDM1-standalone.pdf
}

\end{itemize}

\clearpage
\raggedright



\addcontentsline{toc}{section}{References}

\begin{thebibliography}{100}
\expandafter\ifx\csname url\endcsname\relax
  \def\url#1{\texttt{#1}}\fi
\expandafter\ifx\csname urlprefix\endcsname\relax\def\urlprefix{URL }\fi

\bibitem{LRP2007}
R.~Tribble, et~al., \\
  \href{http://science.energy.gov/~/media/np/nsac/pdf/docs/nuclear_science_low%
_res.pdf}{\tt{http://science.energy.gov/\~/media/np/nsac/pdf/docs/nuclear\_sci%
ence\_low\_res.pdf}} (2007).

\bibitem{HiggsCMS}
T.~C. Collaboration,
  \href{http://arxiv.org/abs/1207.7235}{\tt{http://arxiv.org/abs/1207.7235}}
  (2012).

\bibitem{HiggsATLAS}
T.~A. Collaboration,
  \href{http://arxiv.org/abs/1207.7214}{\tt{http://arxiv.org/abs/1207.7214}}
  (2012).

\bibitem{Sakharov}
A.~D. Sakharov, J. Exp. Theo. Phys. 5 (1967) 24.

\bibitem{Rod03}
V.~A. Rodin, et~al., Phys.\ Rev. C 68 (2003) 044302.

\bibitem{Rod06}
V.~A. Rodin, et~al., Nucl.\ Phys. A 766 (2006) 107, erratum
  \href{http://arxiv.org/pdf/0706.4304v1.pdf}{\tt{http://arxiv.org/pdf/0706.43%
04v1.pdf}}.

\bibitem{Suh05}
J.~Suhonen, Phys. Lett. B 607 (2005) 87.

\bibitem{Kor07}
M.~Kortelainen, et~al., Phys. Lett. B 647 (2007) 128.

\bibitem{Fae08}
A.~Faessler, et~al., Phys. Rev. D 79 (2008) 053001.

\bibitem{Sim09a}
F.~{\protect \v{S}}imkovic, A.~Faessler, P.~Vogel, Phys. Rev. C 79 (2009)
  015502.

\bibitem{Sim09}
F.~{\protect \v{S}}imkovic, et~al., Phys. Rev. C 79 (2009) 055501.

\bibitem{Cau08}
E.~Caurier, F.~Nowacki, A.~Poves, Eur. Phys. J. A 36 (2008) 195.

\bibitem{Cau08a}
E.~Caurier, et~al., Phys. Rev. Lett. 100 (2008) 052503.

\bibitem{Men09}
J.~Men\protect{\'{e}}ndez, et~al., Nucl. Phys. A 818 (2009) 139.

\bibitem{Bare09}
J.~Barea, F.~Iachello, Phys. Rev. C 79 (2009) 044301.

\bibitem{Chan09}
R.~Chandra, et~al., Europhys. Lett. 86 (2009) 32001.

\bibitem{Rod10}
T.~R. Rodr\'{i}guez, G.~Mart\'{i}nez-Pinedo, Phys. Rev. Lett. 105 (2010)
  252503.

\bibitem{Suh08}
J.~Suhonen, O.~Civitarese, Phys. Lett. B 668 (2008) 277.

\bibitem{Men09a}
J.~Men\protect{\'{e}}ndez, et~al., Phys. Rev. C 80 (2009) 048501.

\bibitem{Schif08}
J.~P. Schiffer, et~al., Phys. Rev. Lett. 100 (2008) 112501.

\bibitem{Kay09}
B.~P. Kay, et~al., Phys. Rev. C 79 (2009) 021301(R).

\bibitem{kla06}
H.~V. Klapdor-Kleingrothaus, I.~V. Krivosheina, Mod. Phys. Lett. A 21 (2006)
  1547.

\bibitem{rodejohann_2012}
W.~Rodejohann,
  \href{http://arxiv.org/pdf/1206.2560.pdf}{\tt{http://arxiv.org/pdf/1206.2560%
.pdf}} (2012).

\bibitem{Auger2012}
M.~Auger, et~al., Phys. Rev. Lett. 109 (2012) 032505.

\bibitem{Aguayo2011}
E.~Aguayo, et~al.,
  \href{http://arxiv.org/pdf/1109.6913.pdf}{\tt{http://arxiv.org/pdf/1109.6913%
.pdf}} (2011).

\bibitem{Phillips2011}
D.~G.~{\protect II}. Phillips, et~al.,
  \href{http://arxiv.org/pdf/1111.5578.pdf}{\tt{http://arxiv.org/pdf/1111.5578%
.pdf}} (2011).

\bibitem{Schubert2011}
A.~G. Schubert, et~al.,
  \href{http://arxiv.org/pdf/1109.1567.pdf}{\tt{http://arxiv.org/pdf/1109.1567%
.pdf}} (2011).

\bibitem{Aless11}
F.~Alessandria, et~al.,
  \href{http://arxiv.org/pdf/1109.0494.pdf}{\tt{http://arxiv.org/pdf/1109.0494%
.pdf}}, Submitted to Astropart. Phys. (2011).

\bibitem{And11}
E.~Andreotti, et~al., Astropart. Phys. 34 (2011) 822.

\bibitem{Gando2012}
A.~Gando, et~al., Phys. Rev. C 85 (2012) 045504.

\bibitem{KLZ_new}
A.~Gando, et~al.,
  \href{http://arxiv.org/pdf/1211.3863.pdf}{\tt{http://arxiv.org/pdf/1211.3863%
.pdf}} (2012).

\bibitem{che05}
M.~C. Chen, Nucl. Phys. Proc. Suppl. 145 (2005) 65.

\bibitem{Chen2008}
M.~C. Chen, et~al., eConf (2008)
  C080730\href{http://arxiv.org/ftp/arxiv/papers/0810/0810.3694.pdf}{\tt{http:%
//arxiv.org/ftp/arxiv/papers/0810/0810.3694.pdf}}.

\bibitem{Gom11}
E.~G{\protect\'{o}}mez, et~al.,
  \href{http://arxiv.org/pdf/1106.3630.pdf}{\tt{http://arxiv.org/pdf/1106.3630%
.pdf}} (2011).

\bibitem{Yah10}
N.~Yahlali, et~al., Nucl. Instrum. Meth. Phys. A 617 (2010) 520.

\bibitem{Arn10}
R.~Arnold, et~al., Eur. Phys. J. C 70 (2010) 927.

\bibitem{SUPERNEMO_new}
F.~Piquemal, talk presented at Neutrino 2012, the XXV International COnference
  on Neutrino Physics and
  Astrophysics\href{http://neu2012.kek.jp/index.html}{\tt{http://neu2012.kek.j%
p/index.html}} (2012).

\bibitem{Kumar2011}
K.~Kumar, et~al., \\
  \href{http://science.energy.gov/~/media/np/nsac/pdf/docs/NSAC_NeutronReport.%
pdf}{\tt{http://science.energy.gov/$\sim$/media/np/nsac/pdf/docs/NSACNeutronRe%
port.pdf}} (2011).

\bibitem{Griffith09}
W.~C. Griffith, et~al., Phys. Rev. Lett. 102 (2009) 101601.

\bibitem{Baker06}
C.~A. Baker, et~al., Phys. Rev. Lett. 97 (2006) 131801.

\bibitem{Hudson11}
J.~J. Hudson, et~al., Nature 473 (2011) 493.

\bibitem{De09}
S.~De, et~al., Phys. Rev. A 79 (2009) 041402.

\bibitem{Guest07}
J.~R. Guest, et~al., Phys. Rev. Lett. 98 (2007) 093001.

\bibitem{edmweb}
Proposal available at
  \href{http://www.bnl.gov/edm}{\tt{http://www.bnl.gov/edm}}.

\bibitem{GolubLamoreaux}
R.~Golub, S.~K. Lamoreaux, Phys. Rep. 237 (1994) 1.

\bibitem{Abe08}
S.~Abe, et~al., Phys. Rev. Lett. 100 (2008) 221803.

\bibitem{Matsubara2012}
T.~Matsubara, et~al.,
  \href{http://arxiv.org/pdf/1205.6685.pdf}{\tt{http://arxiv.org/pdf/1205.6685%
.pdf}} (2012).

\bibitem{An2012}
F.~P. An, et~al., Phys. Rev. Lett. 108 (2012) 171803.

\bibitem{Ahn2012}
J.~K. Ahn,
  \href{http://arxiv.org/pdf/1204.0626.pdf}{\tt{http://arxiv.org/pdf/1204.0626%
.pdf}} (2012).

\bibitem{NISTNCR}
\href{http://www.ncnr.nist.gov/}{\tt{http://www.ncnr.nist.gov/}}.

\bibitem{HFIR}
\href{http://neutrons.ornl.gov/facilities/HFIR/}{\tt{http://neutrons.ornl.gov/%
facilities/HFIR/}}.

\bibitem{ATR}
\href{http://atrnsuf.inl.gov}{\tt {http://atrnsuf.inl.gov}}.

\bibitem{Aguilar2009}
A.~A. Aguilar-Arevalo, et~al., Phys. Rev. Lett. 102 (2009) 101802.

\bibitem{Aguilar2010}
A.~A. Aguilar-Arevalo, et~al., Phys. Rev. Lett. 105 (2010) 181801.

\bibitem{agu01}
A.~Aguilar, et~al., Phys. Rev. D 64 (2001) 112007.

\bibitem{Abazajian2012}
K.~Abazajian, et~al.,
  \href{http://arxiv.org/pdf/1204.5379.pdf}{\tt{http://arxiv.org/pdf/1204.5379%
.pdf}} (2012).

\bibitem{Avignone:2003ep}
F.~T. Avignone, Y.~V. Efremenko, J. Phys. G 29 (2003) 2615.

\bibitem{nusns}
\href{http://www.phy.ornl.gov/nusns/}{\tt{http://www.phy.ornl.gov/nusns/}}.

\bibitem{snsworkshop}
\href{http://www.phy.duke.edu/~schol/sns_workshop}{\tt{http://www.phy.duke.edu%
/$\sim$schol/sns$\_$workshop}}.

\bibitem{Efremenko:2008an}
Y.~V. Efremenko, W.~R. Hix, J. Phys. Conf. Ser. 173 (2009) 012006.

\bibitem{Robertson2008}
R.~G.~H. Robertson, J. Phys. Conf. Ser. 120 (2008) 052028,
  \href{http://arxiv.org/pdf/0712.3893.pdf}{\tt{http://arxiv.org/pdf/0712.3893%
.pdf}}.

\bibitem{Monreal2009}
B.~Monreal, J.~Formaggio, Phys. Rev. D 80 (2009) 051301.

\bibitem{Gatti2008}
F.~Gatti, et~al., J. Low Temp. Phys. 151 (2008) 603.

\bibitem{Andreotti2007}
E.~Andreotti, et~al., Nucl. Instrum. Meth. Phys. Res. A 572 (2007) 208.

\bibitem{Hannestad2006}
S.~Hannestad, H.~Tu, Y.~Y.~Y. Wong, JCAP 0606 (2006) 025.

\bibitem{Lesgourgues2006}
J.~Lesgourgues, et~al., Phys. Rev. D 73 (2006) 045021.

\bibitem{Cirigliano:2009bz}
V.~Cirigliano, R.~Kitano, Y.~Okada, P.~Tuzon, Phys. Rev. D 80 (2009) 013002.

\bibitem{mu2e}
\href{http://mu2e.fnal.gov}{\tt{http://mu2e.fnal.gov}}.

\bibitem{Holdom:1985ag}
B.~Holdom, Phys.Lett. B 166 (1986) 196.

\bibitem{Galison:1983pa}
P.~Galison, A.~Manohar, Phys.Lett. B 136 (1984) 279.

\bibitem{Fayet:2007ua}
P.~Fayet, Phys. Rev. D 75 (2007) 115017.

\bibitem{ArkaniHamed:2008qp}
N.~Arkani-Hamed, N.~Weiner, JHEP 0812 (2008) 104.

\bibitem{Goodsell:2009xc}
M.~Goodsell, J.~Jaeckel, J.~Redondo, A.~Ringwald, JHEP 0911 (2009) 027.

\bibitem{Cheung:2009qd}
C.~Cheung, J.~T. Ruderman, L.-T. Wang, I.~Yavin, Phys. Rev. D 80 (2009) 035008.

\bibitem{Morrissey:2009ur}
D.~E. Morrissey, D.~Poland, K.~M. Zurek, JHEP 07 (2009) 050.

\bibitem{Jaeckel:2010ni}
J.~Jaeckel, A.~Ringwald, Ann. Rev. Nucl. Part. Sci. 60 (2010) 405.

\bibitem{Pospelov:2008zw}
M.~Pospelov, Phys. Rev. D 80 (2009) 095002.

\bibitem{ArkaniHamed:2008qn}
N.~Arkani-Hamed, D.~P. Finkbeiner, T.~R. Slatyer, N.~Weiner, Phys. Rev. D 79
  (2009) 015014.

\bibitem{Pospelov:2008jd}
M.~Pospelov, A.~Ritz, Phys. Lett. B 671 (2009) 391.

\bibitem{Essig:2009nc}
R.~Essig, P.~Schuster, N.~Toro, Phys. Rev. D 80 (2009) 015003.

\bibitem{Reece:2009un}
M.~Reece, L.-T. Wang, JHEP 07 (2009) 051.

\bibitem{Strassler:2006im}
M.~J. Strassler, K.~M. Zurek, Phys. Lett. B 651 (2007) 374.

\bibitem{aubert:2009cp}
B.~Aubert, et~al., Phys. Rev. Lett. 103 (2009) 081803.

\bibitem{Batell:2009yf}
B.~Batell, M.~Pospelov, A.~Ritz, Phys. Rev. D 79 (2009) 115008.

\bibitem{Baumgart:2009tn}
M.~Baumgart, C.~Cheung, J.~T. Ruderman, L.-T. Wang, I.~Yavin, JHEP 04 (2009)
  014.

\bibitem{Bjorken:2009mm}
J.~D. Bjorken, R.~Essig, P.~Schuster, N.~Toro, Phys. Rev. D 80 (2009) 075018.

\bibitem{Abrahamyan:2011gv}
S.~Abrahamyan, et~al., Phys. Rev. Lett. 107 (2011) 191804.

\bibitem{Merkel:2011ze}
H.~Merkel, et~al., Phys. Rev. Lett. 106 (2011) 251802.

\bibitem{Essig:2010xa}
R.~Essig, P.~Schuster, N.~Toro, B.~Wojtsekhowski, JHEP 1102 (2011) 009.

\bibitem{HPS}
{The Heavy Photon Search Collaboration (HPS)},
  \href{https://confluence.slac.stanford.edu/display/hpsg/}{https://confluence%
.slac.stanford.edu/display/hpsg/}.

\bibitem{Freytsis:2009bh}
M.~Freytsis, G.~Ovanesyan, J.~Thaler, JHEP 01 (2010) 111.

\bibitem{Wojtsekhowski:2012zq}
B.~Wojtsekhowski, D.~Nikolenko, I.~Rachek,
  \href{http://arxiv.org/pdf/1207.5089.pdf}{\tt{http://arxiv.org/pdf/1207.5089%
.pdf}} (2012).

\bibitem{Batell:2009di}
B.~Batell, M.~Pospelov, A.~Ritz, Phys.Rev. D 80 (2009) 095024.

\bibitem{Essig:2010gu}
R.~Essig, R.~Harnik, J.~Kaplan, N.~Toro, Phys.Rev. D 82 (2010) 113008.

\bibitem{Andreas:2012mt}
S.~Andreas, C.~Niebuhr, R.~Andreas,
  \href{http://arxiv.org/pdf/1209.6083.pdf}{\tt{http://arxiv.org/pdf/1209.6083%
.pdf}} (2012).

\bibitem{e2tau}
M.~Gonderinger, M.~J. Ramsey-Musolf,
  \href{http://arxiv.org/abs/1006.5063}{\tt{http://arxiv.org/abs/1006.5063}}
  (2010).

\bibitem{g-2}
G.~W. Bennett, et~al., Phys. Rev. Lett. 92 (2004) 161802.

\bibitem{Bhattacharya:2011qm}
T.~Bhattacharya, V.~Cirigliano, S.~D. Cohen, A.~Filipuzzi, M.~Gonzalez-Alonso,
  et~al., Phys. Rev. D 85 (2012) 054512.

\bibitem{Bauman:2012fx}
S.~Bauman, J.~Erler, M.~Ramsey-Musolf,
  \href{http://arxiv.org/pdf/1204.0035.pdf}{\tt{http://arxiv.org/pdf/1204.0035%
.pdf}} (2012).

\bibitem{Cirigliano:2012ab}
V.~Cirigliano, M.~Gonzalez-Alonso, M.~L. Graesser,
  \href{http://arxiv.org/pdf/1210.4553.pdf}{\tt{http://arxiv.org/pdf/1210.4553%
.pdf}} (2012).

\bibitem{DubbersSchmidt2011}
D.~Dubbers, M.~G. Schmidt, Rev. Mod. Phys. 83 (2011) 1111.

\bibitem{WietfeldtGreene2011}
F.~E. Wietfeldt, G.~L. Greene, Rev. Mod. Phys. 83 (2011) 1173.

\bibitem{PDG2012}
J.~Beringer, et~al., Phys. Rev. D 86  010001, (Particle Data Group).

\bibitem{Liu2011}
J.~Liu, et~al., Phys. Rev. Lett. 105 (2011) 181803.

\bibitem{Plaster2012}
B.~Plaster, et~al., Phys. Rev. C 86 (2012) 055501.

\bibitem{Wietfeldt2009}
F.~E. Wietfeldt, et~al., Nucl. Instrum. Meth. A 83 (2009) 611.

\bibitem{Pocanic2009}
D.~Pocanic, et~al., Nucl. Instrum. Meth. A 611 (2009) 211.

\bibitem{Bowman2005}
J.~D. Bowman, J. Res. Natl. Inst. Stand. Technol. 110 (2005) 407.

\bibitem{Bernard2004}
V.~Bernard, et~al., Phys. Lett. B 593 (2004) 105, erratum, {\it ibid.} B 599
  (2004) 348.

\bibitem{GardnerHe2012}
S.~Gardner, D.~He, Phys. Rev. D 86 (2012) 016003.

\bibitem{Drees03}
M.~Drees, M.~Rauch, Eur. Phys. J. C 29 (2003) 573.

\bibitem{Mumm2011}
H.~P. Mumm, et~al., Phys. Rev. Lett. 107 (2011) 102301.

\bibitem{Cooper12}
R.~L. Cooper, et~al., Nucl. Instrum. Meth. A 691 (2012) 64.

\bibitem{piref_1a}
D.~Po\v{c}ani\'c, et~al., Proc. Sci. CD09 (2009) 009.

\bibitem{piref_1b}
\href{http://pen.phys.virginia.edu}{\tt{http://pen.phys.virginia.edu}}.

\bibitem{piref_2a}
M.~Aoki, et~al., Phys. Rev. D 84 (2011) 052002.

\bibitem{piref_2b}
\href{http://pienu.triumf.ca}{\tt{http://pienu.triumf.ca}}.

\bibitem{piref_3a}
W.~J. Marciano, A.~Sirlin, Phys. Rev. Lett 71 (1993) 3629.

\bibitem{piref_3b}
M.~Finkemeier, Phys. Lett. B 387 (1996) 391.

\bibitem{Cirigliano:2007xi}
V.~Cirigliano, I.~Rosell, Phys. Rev. Lett. 99 (2007) 231801.

\bibitem{PVES1}
M.~J. Ramsey-Musolf, Phys. Rev. C 60 (1999) 015501.

\bibitem{PVES2}
R.~D. Young, R.~Carlini, A.~Thomas, J.~Roche, Phys. Rev. Lett. 99 (2007)
  122003.

\bibitem{PVES3}
R.~Carlini, et~al., the QWeak Experiment, Jefferson Lab Experiment E12-08-016,
  \href{http://arxiv.org/abs/1202.1255}{\tt{http://arxiv.org/abs/1202.1255}}.

\bibitem{PVES4}
K.~Kumar, et~al., the MOLLER Experiment, Jefferson Lab Experiment E12-09-005,
  \href{http://hallaweb.jlab.org/12GeV/Moller/}{\tt{http://hallaweb.jlab.org/1%
2GeV/Moller/}}.

\bibitem{PVES5}
P.~A. Souder, et~al., the SoLID Experiment, Jefferson Lab Experiment
  E12-10-007,
  \href{http://hallaweb.jlab.org/parity/PR-10-007-SoLID-PVDIS.pdf}{\tt{http://%
hallaweb.jlab.org/parity/PR-10-007-SoLID-PVDIS.pdf}}.

\bibitem{Bueno11}
J.~F. Bueno, et~al., Phys. Rev. D 84 (2011) 032005.

\bibitem{Bayes11}
R.~Bayes, et~al., Phys. Rev. Lett. 106 (2011) 041804.

\bibitem{Webber11}
D.~M. Webber, et~al., Phys. Rev. Lett. 106 (2011) 041803.

\bibitem{mucap1}
V.~A. Andreev, et~al., Phys. Rev. Lett. 99 (2007) 032002.

\bibitem{mucap2}
V.~A. Andreev, et~al.,
  \href{http://arxiv.org/pdf/1210.6545v1}{\tt{http://arxiv.org/pdf/1210.6545v1%
}}, submitted to Phys. Rev. Lett. (2012).

\bibitem{MuCAP_THEORY}
A.~Czarnecki, W.~J. Marciano, A.~Sirlin, Phys. Rev. Lett. 99 (2007) 032003.

\bibitem{MUSUN}
V.~A. Andreev, et~al.,
  \href{http://arxiv.org/pdf/1004.1754v1}{\tt{http://arxiv.org/pdf/1004.1754v1%
}} (2010).

\bibitem{g-2_redux}
\href{http://gm2.fnal.gov}{\tt{http://gm2.fnal.gov}}.

\bibitem{Gericke2011}
M.~Gericke, et~al., Phys. Rev. C 83 (2011) 015505.

\bibitem{Bowman2007}
J.~D. Bowman, et~al., proposal Submitted to the SNS FNPB.

\bibitem{Snow2011}
W.~M. Snow, et~al., Phys. Rev. C 83 (2011) 022501(R).

\bibitem{Abrahamyan:2012gp}
S.~Abrahamyan, et~al., Phys. Rev. Lett. 108 (2012) 112502.

\bibitem{Tsang:2012se}
M.~B. Tsang, et~al., Phys. Rev. C 86 (2012) 015803.

\bibitem{Bellini2011}
G.~Bellini, et~al., Phys. Rev. Lett. 107 (2011) 141302.

\bibitem{Bellini2012}
G.~Bellini, et~al.,
  \href{http://arxiv.org/pdf/1110.3230.pdf}{\tt{http://arxiv.org/pdf/1110.3230%
.pdf}} (2012).

\bibitem{geo_neutrinos}
S.~Dye, Rev. Geophys. 50 (2012) RG3007,
  \href{http://arxiv.org/pdf/1111.6099.pdf}{\tt{http://arxiv.org/pdf/1111.6099%
.pdf}}.

\bibitem{ScholSN}
K.~Scholberg,
  \href{http://arxiv.org/pdf/1205.6003v1}{\tt{http://arxiv.org/pdf/1205.6003v1%
}} (2012).

\bibitem{Scholberg:2005qs}
K.~Scholberg, Phys. Rev. D 73 (2006) 033005.

\bibitem{Scholberg:2009ha}
K.~Scholberg, et~al.,
  \href{http://arxiv.org/pdf/0910.1989.pdf}{\tt{http://arxiv.org/pdf/0910.1989%
.pdf}} (2009).

\bibitem{reactor_nu_monitors}
A.~Bernstein, et~al., Science and Global Security 18 (2010) 127,
  \href{http://arxiv.org/abs/0908.4338}{\tt{http://arxiv.org/abs/0908.4338}}.

\end{thebibliography}
\end{document}